%% file: main.tex
\documentclass[10pt,journal]{IEEEtran}

\usepackage{cite}
\usepackage{amsmath,amssymb,bm}
\usepackage{graphicx}
\usepackage{booktabs}
\usepackage{multirow}
\usepackage{tabularx}
\usepackage{float}
\usepackage[caption=false,font=footnotesize]{subfig}
\usepackage{url}
\usepackage[hidelinks]{hyperref}
\hypersetup{
  pdftitle={Single-Model Adaptive Wireless Image Transmission via Feature Sparsity Regularization},
  pdfauthor={Xianghao Cui, Li Lan, Qi He, Bo Che, Chenyuan Feng, Zhi Chen, and Tony Q. S. Quek},
  pdfkeywords={joint source-channel coding, wireless image transmission, structured sparsity, content-adaptive transmission, SNR adaptation, rate adaptation}
}

\graphicspath{{figures/}}

\newcommand{\cbr}{\ensuremath{\mathrm{CBR}}}
\newcommand{\snr}{\ensuremath{\mathrm{SNR}}}
\newcommand{\awgn}{\ensuremath{\mathrm{AWGN}}}
\newcommand{\rayleigh}{\ensuremath{\mathrm{Rayleigh}}}
\newcommand{\pn}{\ensuremath{\mathrm{PN}}}

\title{Single-Model Adaptive Wireless Image Transmission via Feature Sparsity Regularization}

\author{Xianghao~Cui, Li~Lan, Qi~He,~\IEEEmembership{Member,~IEEE,} Bo~Che, Chenyuan~Feng,~\IEEEmembership{Member,~IEEE,} 

Zhi~Chen,~\IEEEmembership{Senior Member,~IEEE,} and Tony Q. S. Quek,~\IEEEmembership{Fellow,~IEEE}%
\thanks{Xianghao Cui, Li Lan, Qi He, Bo Che, and Zhi Chen are with the National Key Laboratory of Wireless Communications, University of Electronic Science and Technology of China, Chengdu 611731, China.
Corresponding author: Qi He (e-mail: heqi@uestc.edu.cn).

Chenyuan Feng is with the Department of Computer Science, University of Exeter, Exeter EX4 4QF, U.K.

Tony Q. S. Quek is with the Singapore University of Technology and Design, Singapore 487372.
}%
}

\begin{document}

\maketitle

\begin{abstract}
Learned joint source--channel coding (JSCC) enables robust wireless image transmission by jointly optimizing the transmitter and receiver over differentiable channel models. For bandwidth-limited and time-varying visual links, a single model should support user-adjustable transmission rate and adapt to changing wireless channel conditions, while also dynamically allocating resources according to spatial content. Existing content-adaptive or dynamic allocation schemes often rely on entropy coding, context/probability prediction, explicit rate maps or masks, or auxiliary allocation networks, complicating the encoder--decoder pipeline and increasing side-information overhead. We propose TS-JSCC, a single-model adaptive JSCC framework with tail-structured sparsification. First, an L1-based tail-structured sparsification objective encourages each token to retain an active feature-channel prefix while suppressing trailing ones. This enables content-adaptive feature-channel allocation with compact side information through active-prefix transmission. Second, lightweight stage-wise neural regulating modules use a normalized sparsity-control coefficient and the channel signal-to-noise ratio (SNR) to rescale intermediate features for single-model transmission rate and SNR adaptation. Experiments on CIFAR-10, Kodak, and CLIC2021 under additive white Gaussian noise (AWGN) and Rayleigh fading show that TS-JSCC achieves 
strong rate--distortion performance against the latest learned-JSCC baselines and remains competitive with the considered idealized separation baselines, while retaining a simple one-shot encoder--decoder without extra structures or computations. 
\end{abstract}


\section{Introduction}

Wireless image transmission is a core component of edge intelligence, immersive communication, remote sensing, connected vehicles, and low-latency visual telemetry. These applications require images to be delivered over bandwidth-limited and time-varying wireless links while preserving visual quality under strict delay and reliability constraints. Conventional digital pipelines follow Shannon's source--channel separation principle: an image is first compressed into a bitstream by a source codec, ranging from conventional codecs such as JPEG \cite{Wallace1992} and BPG \cite{Bellard2014} to advanced learned neural image-compression models \cite{HPCM2025,He2022ELIC,DCAE2025,LALIC2025,MambaIC2025}, and is then protected by a channel code such as LDPC \cite{RichardsonUrbanke2001} or Polar coding \cite{Arikan2009} before modulation and transmission \cite{Shannon1948,Shannon1959,CoverThomas2006}. Although separation is asymptotically optimal under ideal assumptions, practical visual delivery often faces finite blocklengths, link mismatch, and rapidly varying channels, which can cause cliff effect and saturation effect, degrading reconstruction quality and reducing transmission efficiency \cite{wang2015mcscast,jakubczak2010softcast}.

Learned joint source--channel coding (JSCC) offers an alternative for robust wireless image transmission \cite{Bourtsoulatze2019}. Instead of optimizing source coding and channel coding separately, learned JSCC trains an end-to-end neural encoder--decoder that maps an image directly to continuous-valued wireless channel symbols over a differentiable wireless channel model. This formulation allows visual representation learning, source compression, channel robustness, and image reconstruction to be optimized jointly. Representative developments include convolutional DeepJSCC transmitters \cite{Bourtsoulatze2019} and Swin Transformer-based JSCC backbones \cite{Liu2021Swin,SwinJSCC}.

\begin{figure}[t]
\centering
\includegraphics[width=\linewidth]{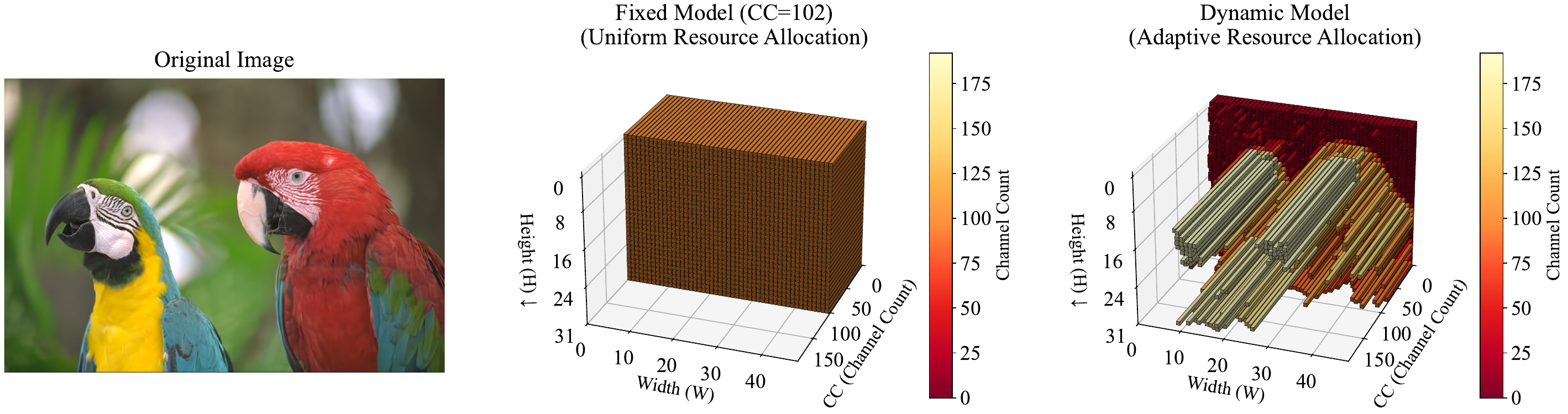}
\caption{\textbf{Visualization of uniform and content-adaptive resource allocation} under the same communication budget. In this paper, transmission rate is measured by the channel bandwidth ratio (CBR). The fixed model is produced by the SwinJSCC baseline~\cite{SwinJSCC} with a constant latent feature-channel dimension \(C=102\) (\(\mathrm{CBR}=0.0664\)), whereas the dynamic model is produced by the proposed TS-JSCC with a larger maximum dimension \(C_{\max}=192\) and activates only a subset of feature channels to satisfy the same average CBR constraint.}
\label{fig:allocation_compare_intro}
\end{figure}

Despite these advances, a key limitation remains: mainstream learned JSCC systems often assign the same number of feature channels to every spatial token, where each token denotes the \(C\)-dimensional feature vector at location \((h,w)\) in an \(H_z\times W_z\times C\) latent feature map, with \(H_z\) and \(W_z\) denoting the latent height and width. This uniform resource allocation is poorly matched to natural images with highly non-uniform visual content. For example, edges, textures, objects, and salient regions usually need more feature resources than smooth backgrounds. Fig.~\ref{fig:allocation_compare_intro} illustrates this motivation: under the same average budget, content-adaptive allocation can assign more feature resources to detail-rich regions. Recent adaptive JSCC schemes pursue content-dependent allocation through auxiliary policy or prediction modules \cite{RateAdaptiveJSCC_YangKim2022,RateAdaptiveJSCC_Zhang2023}, entropy or probability priors for rate adaptation \cite{Chen2023EntropyAware,Dai2022NTSCC,Wang2023ImprovedNTSCC,Zhang2025HJSCC}, and explicit spatial rate maps or prefix masks \cite{Zhang2023VLSCC}. While confirming the value of non-uniform allocation, these approaches typically depend on auxiliary allocation modules or entropy/probability estimates or priors, thereby incurring considerable additional architectural complexity and side-information overhead.

Beyond efficiency, a practical scheme should adapt to different demands and conditions. A high-quality visual stream typically consumes more transmission symbols, whereas a low-latency or low-rate stream requires fewer. Likewise, favorable wireless channel conditions permit more aggressive compression, while poor conditions necessitate more robust representations. Rather than training multiple separate models, it is more desirable to train a single model capable of adapting to different user-specified rate/quality requirements and wireless channel conditions. Existing adaptive JSCC methods provide useful ingredients, such as differentiable gating \cite{RateAdaptiveJSCC_YangKim2022}, image-wise operating-point prediction \cite{RateAdaptiveJSCC_Zhang2023}, adaptive-model training \cite{RateAdaptiveJSCC_Song2023}, and rate/SNR conditioning \cite{SwinJSCC,Bian2023DeepJSCCLpp,Li2025STARJSCC}. However, simultaneously achieving wide-range transmission rate control, multi-SNR robustness, token-level spatial content adaptivity, and low side-information overhead remains challenging.

This paper proposes TS-JSCC, a tail-structured sparsification framework for single-model adaptive wireless image transmission, which realizes adaptive transmission rates by deciding how many feature channels to retain for each token according to user-specified rate requirements, wireless channel conditions, and token-level spatial content. On the one hand, TS-JSCC introduces feature sparsity regularization on encoder outputs so that end-to-end training learns how many feature channels should be retained for different spatial tokens, rate requirements, and channel conditions. Technically, an L1-based tail-structured sparsification loss encourages the encoder latent to suppress low-utility tail entries; after thresholding, the surviving feature channels form a compact active-prefix representation that keeps only the front active feature channels for each token. In particular, a sliding-window weighting strategy restricts the sparsity pressure to a short region near the feature-channel tail of each token, rather than regularizing all feature channels, thereby preserving more feature representation capacity. Furthermore, a magnitude-aware straight-through estimator (mSTE) is proposed to stabilize end-to-end training through thresholding. Together, these designs realize adaptive allocation purely through training, without adding model structures or computation, as quantified in Table~\ref{tab:complexity_summary}.

On the other hand, to support transmission rate and SNR adaptation with one model, TS-JSCC further introduces lightweight stage-wise regulating modules driven by the channel SNR and a user-specified normalized coefficient \(\lambda_{\mathrm{norm}}\) to control the transmission rate.
These modules produce channel- and rate-dependent scaling factors, enabling the encoder to dynamically transform features based on both the prevailing SNR and the user's rate requirement, thereby balancing source and channel coding accordingly. Symmetric deregulating modules at the decoder invert this adaptation, completing the dual-conditioned adjustment.
The full adaptive architecture of TS-JSCC is illustrated in Fig.~\ref{fig:adaptive_modulation_architecture}. Experiments in Sec.~\ref{sec:results_discussion} demonstrate that TS-JSCC achieves strong learned-JSCC rate--distortion performance under the evaluated AWGN and Rayleigh settings and remains competitive with idealized HPCM~\cite{HPCM2025} + LDPC + QAM separation baselines while retaining a simpler one-shot JSCC pipeline (Figs.~\ref{fig:awgn_foundation}--\ref{fig:snr_robustness}). The adaptive TS-JSCC variants preserve strong reconstruction quality while using one model to cover different transmission rates and SNRs, confirming that the regulating modules provide effective adaptation with only marginal overhead (Figs.~\ref{fig:awgn_foundation}--\ref{fig:snr_robustness} and Table~\ref{tab:complexity_summary}). The MS-SSIM and visual results further show that the learned allocation improves structural fidelity and region of interest (ROI) detail recovery rather than only increasing PSNR (Figs.~\ref{fig:msssim_awgn}, \ref{fig:visual_fixcbr}, and~\ref{fig:visual_low}). Finally, the ablation and overhead analyses verify that the L1 tail sparsification, sliding-window weighting, and mSTE are responsible for the gains, which are obtained with compact side information (Fig.~\ref{fig:ablation_awgn} and Table~\ref{tab:sideinfo_overhead_highres}). To facilitate reproducible research, the source code will be released on GitHub upon publication.

Our main contributions are summarized as follows.
\begin{itemize}
\item We develop a unified TS-JSCC framework in which a single end-to-end model jointly supports user-specified transmission-rate control, SNR adaptation, and token-wise content-adaptive resource allocation.
\item To realize spatial content adaptivity, we introduce feature sparsity regularization on encoder latents. Without extra structures or computations, the proposed \(\mathrm{L1}\)-regularized tail-sparsification loss is optimized end-to-end to induce structured active feature-channel prefixes with low side-information overhead and no additional model complexity.
\item We design lightweight stage-wise regulating modules that inject the channel SNR and a user-specified coefficient into the proposed TS-JSCC framework, enabling one model to adapt to different rate requirements and channel conditions with marginal complexity overhead.
\item We validate TS-JSCC on CIFAR-10, Kodak, and CLIC2021 under AWGN and Rayleigh channels. The proposed scheme achieves strong performance against the latest learned-JSCC baselines and remains competitive with the considered separation baseline.
\end{itemize}

The rest of this paper is organized as follows. Section II reviews related work. Section III presents the system model and problem formulation. Section IV introduces the proposed TS-JSCC framework. Section V describes the experimental setup. Section VI reports the results and discussion. Section VII concludes the paper.

\section{Related Work}

\subsection{Adaptive Rate Allocation}

Adaptive rate allocation has been studied in both learned JSCC and learned compression. For clarity, we describe representative methods from coarser to finer allocation levels, while noting that learned-compression allocation is implicit bit allocation rather than direct JSCC symbol selection. In learned JSCC, predictive adaptive coding \cite{RateAdaptiveJSCC_Zhang2023} and adaptive-rate DeepJSCC \cite{RateAdaptiveJSCC_YangKim2022} operate at the \textbf{sample level} by using an auxiliary predictor or policy network to select one operating point or symbol budget for the whole input image. SwinJSCC w/ RA \cite{SwinJSCC} operates at the \textbf{feature-map level}: Rate ModNet generates a spatially shared channel mask, so a selected feature channel is kept or suppressed over the whole latent feature map; entropy-aware adaptive control \cite{Chen2023EntropyAware} also estimates latent importance and activates or prunes feature maps/symbols with entropy guidance. VLSCC \cite{Zhang2023VLSCC}, NTSCC \cite{Dai2022NTSCC} and NTSCC++ \cite{Wang2023ImprovedNTSCC}, and HJSCC \cite{Zhang2025HJSCC} operate at the \textbf{token level}: VLSCC predicts spatial rate maps or prefix masks; NTSCC and NTSCC++ use latent-prior entropy to assign variable bandwidths to semantic embedding vectors; and the most recent HJSCC uses entropy-guided hierarchical prefix decisions with spatial grouping, and thus different spatial tokens or token groups may use different prefix lengths. 

In learned compression, HPCM \cite{HPCM2025} performs \textbf{latent-element-wise} bit allocation in the compression domain: hierarchical context/probability modeling assigns different entropy-coded bit lengths to quantized latent elements conditioned on their spatial contexts before channel coding. This latent-element-wise allocation can be highly flexible, but it depends on probability priors, entropy-coded bitstreams, and successful bitstream recovery. Although these methods improve non-uniform resource use, they often require auxiliary predictors, entropy/probability priors, explicit maps or masks, or multi-level coordination, thereby incurring considerable additional architectural complexity and side-information overhead.

TS-JSCC also performs \textbf{token-level} allocation, but its main distinction lies in the allocation mechanism. It induces adaptive feature-channel allocation through \(\mathrm{L1}\)-regularized tail sparsification rather than through a policy network, entropy/probability prior, context model, or explicit rate map. Consequently, the receiver needs only a compact termination index for each token instead of a dense mask, and the wireless transmission path avoids entropy-coded bitstream recovery. This makes the pipeline more friendly to one-shot parallel inference and practical throughput.

\subsection{SNR- and Rate-Adaptive Learned Transmission}

Practical deployment requires a single model to operate across varying SNR conditions and transmission rates. Along the SNR axis, adaptive deep JSCC conditions feature-wise attention on the instantaneous SNR so that a single network adapts its features to varying wireless channel quality \cite{Xu2022ADJSCC}. The attention-based approach improves channel-state robustness, but it does not provide rate/quality-adaptive transmission rate adjustment or content-adaptive allocation. 
Along the rate axis, adaptive-rate DeepJSCC employs differentiable symbol selection to activate different subsets of candidate channel symbols under different rate budgets \cite{RateAdaptiveJSCC_YangKim2022}. It supports adaptive rate control, but relies on an additional policy/symbol-selection mechanism rather than inducing structured sparsity inside the latent representation selected for transmission. Beyond these separate adaptation mechanisms, joint rate/SNR adaptation has also been explored in DeepJSCC-l++ \cite{Bian2023DeepJSCCLpp}, STARJSCC \cite{Li2025STARJSCC}, and SwinJSCC \cite{SwinJSCC}. Among these joint adaptive methods, SwinJSCC leverages a Swin Transformer backbone and has demonstrated strong reconstruction performance, making it a principal adaptive-JSCC baseline in our experiments. It studies SNR adaptation, rate adaptation, and their joint configuration within its adaptive JSCC framework. Its Rate ModNet can regulate the target CBR, yet the rate decision is mainly a feature-channel mask rather than token-level content-adaptive feature-channel allocation, which limits its ability to concentrate transmission resources on spatially informative regions.

These methods establish the value of conditioning on wireless channel quality and rate, but they do not simultaneously provide rate/quality adaptation, SNR adaptation, token-level content-adaptive allocation, and lightweight model complexity. TS-JSCC instead couples the proposed lightweight SNR/rate regulating modules with per-token feature-channel allocation produced by latent sparsification, so the global operating point and spatially varying symbol budget are learned jointly within the same encoder--decoder. Because the regulating modules are small stage-wise MLPs, their additional parameters and computation are marginal, as quantified in Table~\ref{tab:complexity_summary}.

\subsection{Side Information, Structured Sparsity, and Surrogate Gradients}

For adaptive visual transmission, the selected allocation pattern must be recoverable at the receiver, so the side information required to describe that pattern becomes an important part of the effective communication cost. If the transmitter removes arbitrary symbols or assigns independent rates to spatial locations, this side information usually takes the form of dense masks, spatial rate maps, or multi-level coordination metadata for restoring the latent tensor \cite{Chen2023EntropyAware,Zhang2023VLSCC,Zhang2025HJSCC}. These designs enlarge the allocation space, but the side-information overhead can offset part of the resource saving. Therefore, useful sparsification for JSCC should not only remove redundant symbols, but also impose a support geometry that can be represented compactly as side information.

Structured sparsity addresses this side-information issue by restricting the support to regular patterns rather than arbitrary masks. Classical sparse regularizers such as the lasso and Group Lasso \cite{Tibshirani1996,YuanLin2006GroupLasso}, structured pruning and learnable gates \cite{WenSSL,Han2015Pruning,HeChannelPruning,LiuNetworkSlimming}, and stochastic sparsification methods based on \(L_0\) or variational regularization \cite{Louizos2018L0,Molchanov2017VariationalDropout} all encourage structured active sets. In contrast to model pruning, TS-JSCC applies sparsity to latent symbols selected for transmission and uses a tail-contiguous support along the feature-channel dimension, so that each token support reduces to one termination index. Realizing such a support, however, requires selecting an active feature-channel prefix per token, which is naturally expressed through hard thresholding on feature-channel magnitudes followed by prefix formation, two discrete operations that are not directly differentiable.

Training through these discrete operations therefore requires gradient surrogates. Straight-through estimators, Gumbel-Softmax, Concrete relaxations, and surrogate-gradient analyses provide standard tools for discrete or nearly discrete operations \cite{Bengio2013STE,Jang2017GumbelSoftmax,Maddison2017Concrete,Yin2019UnderstandingSTE}. TS-JSCC adopts this perspective for communication resource allocation and introduces a magnitude-aware straight-through estimator (mSTE) to train active-prefix transmission end to end.

\section{System Model and Problem Formulation}

\begin{figure}[t]
\centering
\includegraphics[width=\linewidth]{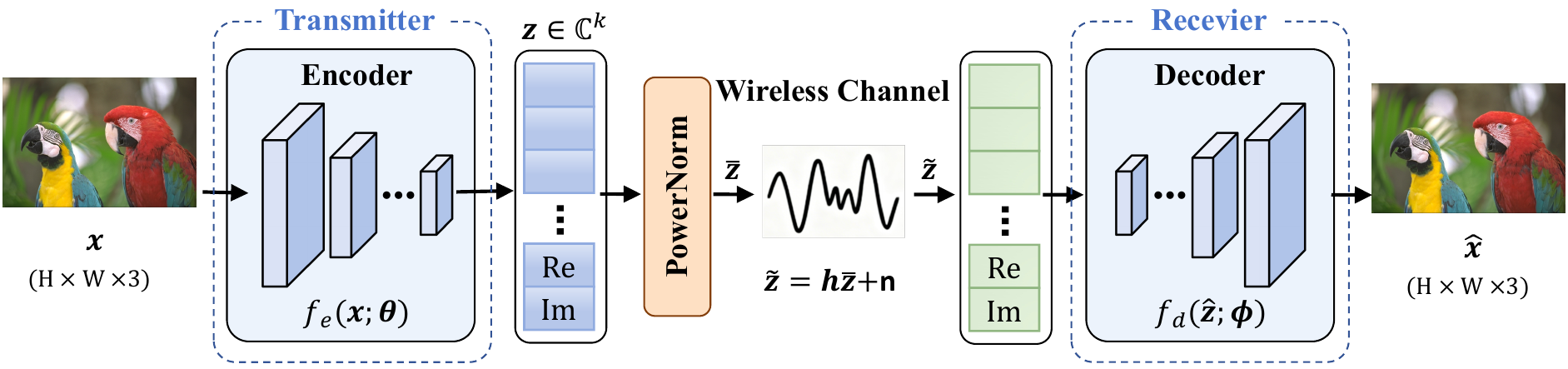}
\caption{\textbf{Learned JSCC system overview.}}
\label{fig:system_chain}
\end{figure}

\subsection{Preliminaries: learned JSCC}

The basic learned JSCC chain is shown in Fig.~\ref{fig:system_chain}. Given an RGB image \(\mathbf{x}\in\mathbb{R}^{H\times W\times 3}\), a JSCC encoder \(f_e(\cdot;\boldsymbol{\theta})\) produces a real-valued latent tensor
\begin{equation}
\mathbf{z}_{\mathbb{R}} = f_e(\mathbf{x};\boldsymbol{\theta}) \in \mathbb{R}^{S\times C},
\end{equation}
where \(S=H_zW_z\) is the number of latent spatial tokens, and \(H_z\) and \(W_z\) denote the height and width of the latent feature map, respectively. For the four-stage high-resolution backbone shown in Fig.~\ref{fig:adaptive_modulation_architecture}, \(H_z=H/16\) and \(W_z=W/16\). \(C\) is the maximum number of \emph{feature channels} per token. By pairing every two real entries into one in-phase/quadrature (I/Q) symbol, we obtain the equivalent complex baseband tensor \(\mathbf{z}\in\mathbb{C}^{S\times (C/2)}\). To avoid ambiguity, we use \emph{feature channel} for the latent dimension \(C\), and \emph{wireless channel} for the physical propagation medium such as \awgn{} or \rayleigh{} fading. Let \(k=SC/2\) denote the number of complex symbols. Before entering the wireless channel, the symbol tensor is power normalized as
\begin{equation}
\bar{\mathbf{z}}
=
\pn(\mathbf{z})
=
\sqrt{\frac{kP}{\|\mathbf{z}\|_F^2}}\,
\mathbf{z},
\label{eq:pn_prelim}
\end{equation}
where \(P\) is the target average transmit power and \(\|\cdot\|_F\) denotes the Frobenius norm. Under an additive white Gaussian noise (AWGN) channel, the received vector is
\begin{equation}
\tilde{\mathbf{z}}=\bar{\mathbf{z}}+\mathbf{n},\qquad \mathbf{n}\sim\mathcal{CN}(\mathbf{0},\sigma^2\mathbf{I}).
\end{equation}
After power normalization, the wireless channel quality is characterized by the signal-to-noise ratio \(\snr=P/\sigma^2\).
For Rayleigh fading, the received signal is
\begin{equation}
\tilde{\mathbf{z}}=\mathbf{h}\odot\bar{\mathbf{z}}+\mathbf{n},
\qquad
\mathbf{h}\sim\mathcal{CN}(\mathbf{0},\mathbf{I}),
\end{equation}
where \(\odot\) denotes element-wise multiplication, \(\mathbf{h}\) is the vector of fading coefficients for the complex channel-input symbols, and \(\mathbf{n}\) is complex Gaussian noise. The decoder \(f_d(\cdot;\boldsymbol{\phi})\) reconstructs the image as \(\hat{\mathbf{x}}=f_d(\tilde{\mathbf{z}};\boldsymbol{\phi})\).

The channel bandwidth ratio (CBR) is measured by the number of sent \emph{complex} symbols per source scalar:
\begin{equation}
\cbr=\frac{k}{N_s},\qquad N_s=3HW.
\end{equation}

A learned JSCC system is trained end to end by minimizing the expected reconstruction distortion over source images and wireless channel realizations:
\begin{equation}
(\boldsymbol{\theta}^*,\boldsymbol{\phi}^*)
=
\arg\min_{\boldsymbol{\theta},\boldsymbol{\phi}}
\mathbb{E}_{\mathbf{x},\,\mathbf{n}}
\big[d(\mathbf{x},\hat{\mathbf{x}})\big].
\label{eq:deepjscc_obj}
\end{equation}

\subsection{Why uniform token lengths are suboptimal}

Mainstream learned JSCC models use the same feature-channel dimension \(C\) for all tokens, which means that all spatial locations consume the same communication budget. This is inefficient because natural images are highly heterogeneous. Regions with edges, object boundaries, fine texture, and semantic foreground content usually require more transmission resources than smooth regions. Under a fixed average budget, assigning the same number of feature-channel symbols to every token wastes bandwidth on redundant content and under-serves the regions that dominate visual fidelity. This imbalance between spatial content complexity and allocated feature-channel resources is later corroborated quantitatively by the rate--distortion curves in Fig.~\ref{fig:awgn_foundation} and Fig.~\ref{fig:rayleigh_cbr}, and qualitatively by the allocation visualizations in Fig.~\ref{fig:visual_fixcbr} and Fig.~\ref{fig:visual_low}.

This observation motivates token-wise truncation along the feature-channel dimension. However, the truncation pattern must be chosen carefully. Arbitrary sparse removal would require dense masks or explicit index maps to recover the latent structure at the receiver, which introduces non-negligible side-information overhead. Our design target is therefore a tail-contiguous structure in which the active feature-channel symbols of each token form an active prefix, i.e., the retained front segment along the feature-channel dimension, and the discarded feature-channel symbols form a suffix. This allows each token to be represented by a single termination index rather than by a dense binary pattern.

\section{Proposed Method}

\begin{figure*}[t]
\centering
\includegraphics[width=\textwidth]{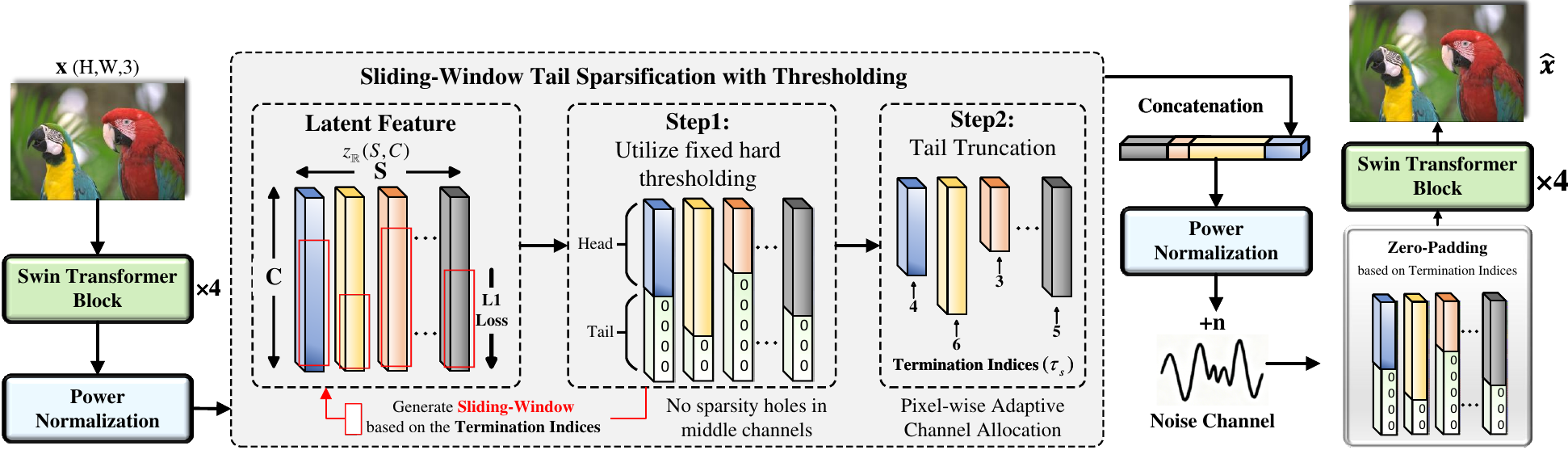}
\caption{\textbf{Tail-sparsification training for content-adaptive allocation.} The tail-sparsification module is inserted between the encoder and wireless channel to impose L1-driven tail-contiguous sparsity, apply thresholding, and form token-wise active feature-channel prefixes.}
\label{fig:overview_framework}
\end{figure*}

\subsection{Adaptive tail sparsification for dynamic resource allocation}
\label{sec:tsjscc}

As illustrated in Fig.~\ref{fig:overview_framework}, we insert the proposed tail-sparsification module between the encoder and the wireless channel. Let \(\mathbf{z}_{\mathbb{R}}=f_e(\mathbf{x};\boldsymbol{\theta})\in\mathbb{R}^{S\times C}\) denote the real encoder latent, and let \(\mathbf{z}\in\mathbb{C}^{S\times (C/2)}\) be its complex I/Q form obtained by pairing the encoder output into in-phase and quadrature components. The tail-sparsification module does not predict token lengths as a separate decision. Instead, it applies tail-structured sparsification and thresholding along the feature-channel dimension; the last active feature-channel position of each token then defines the termination index used to pack the active feature-channel prefixes.

Given a fixed threshold \(\epsilon\), entries with \(|z_{s,c}|<\epsilon\) are discarded from transmission after tail sparsification. For each token \(s\), the termination index is computed as the last active feature-channel position:
\begin{equation}
\tau_s=\max\{c\in\{1,2,\ldots,C/2\}: |z_{s,c}|\ge \epsilon\}.
\label{eq:tau}
\end{equation}
If no entry satisfies \(|z_{s,c}|\ge \epsilon\), then \(\tau_s=0\) by definition. Only the active feature-channel prefix \(\{z_{s,c}\}_{c=1}^{\tau_s}\) is selected for transmission, whereas the suffix \(\{z_{s,c}\}_{c=\tau_s+1}^{C/2}\) is discarded. We write the packed transmit vector as
\begin{equation}
\mathbf{z}_{\mathrm{tx}}
=
\mathrm{pack}\!\left(
\left\{\left[z_{s,1},\ldots,z_{s,\tau_s}\right]\right\}_{s=1}^{S}
\right)
\in\mathbb{C}^{k_{\mathrm{tx}}},
\label{eq:z_tx_pack}
\end{equation}
where \(\mathrm{pack}(\cdot)\) concatenates all active feature-channel prefixes while preserving the token order. The effective number of sent complex symbols is
\begin{equation}
k_{\mathrm{tx}}=\sum_{s=1}^{S}\tau_s \le S(C/2)
\label{eq:k_tx}
\end{equation}
and the corresponding CBR is \(k_{\mathrm{tx}}/N_s\).

In both training and inference, wireless channel corruption is applied only to the active channel-input symbols. The discarded tail is represented by zero padding at the receiver and does not receive noise perturbation.

At the receiver, the original tensor shape is restored by zero-padding according to \(\boldsymbol{\tau}=\{\tau_s\}_{s=1}^{S}\). Hence \(\boldsymbol{\tau}\) must be available as lossless side information. To encode each \(\tau_s\in\{0,1,\ldots,C/2\}\), we use
\begin{equation}
b_{\tau}=\left\lceil \log_2\left(\frac{C}{2}+1\right)\right\rceil
\end{equation}
bits per token, which yields \(B_{\tau}=Sb_{\tau}\) bits per image. Using the AWGN channel-capacity upper bound as a convenient equivalent-rate measure, we convert this side-information burden into an equivalent CBR increase:
\begin{equation}
\Delta\mathrm{CBR}_{\tau}=\frac{B_{\tau}}{N_s\log_2(1+\snr)}.
\end{equation}
This compact termination-index side information is a central advantage of the proposed tail structure over dense masking: the side-information load scales with one integer per token instead of one binary indicator per symbol. As quantified later in Table~\ref{tab:sideinfo_overhead_highres}, the resulting additional CBR is very small in the high-resolution setting. To further reduce side-information signaling, an inference-only 16-state quantizer maps each \(\tau_s\) to \(\hat{\tau}_s\) and signals it with 4 bits per token (Supplementary Section~IV).

As shown in Fig.~\ref{fig:overview_framework}, two power-normalization steps are included around the tail-sparsification operation. The first normalization is applied before the L1 sparsity loss and thresholding. Without this scale constraint, the L1 penalty could shrink the constrained latent entries through global rescaling, making their average power much smaller than the intended transmit-power scale; in that case, the same numerical threshold \(\epsilon\) would no longer correspond to a consistent signal-amplitude criterion. Normalizing the latent first makes the sparsity penalty operate under a consistent power scale, so \(\epsilon\) has a stable physical meaning; for example, \(\epsilon=10^{-2}\) removes only very weak latent symbols relative to the normalized signal scale, which are likely to be dominated by channel noise. After masking and truncation, discarding many low-amplitude symbols changes the average power per active symbol because the symbol count decreases much more than the retained signal energy. Therefore, a second normalization maps the packed active feature-channel-prefix sequence \(\mathbf{z}_{\mathrm{tx}}\) to the channel-input sequence \(\bar{\mathbf{z}}_{\mathrm{tx}}\), enforcing the power constraint over the active channel-input symbols:
\begin{equation}
\bar{\mathbf{z}}_{\mathrm{tx}}
=
\pn(\mathbf{z}_{\mathrm{tx}})
=
\sqrt{\frac{k_{\mathrm{tx}}P}{\|\mathbf{z}_{\mathrm{tx}}\|_2^2}}\,
\mathbf{z}_{\mathrm{tx}}.
\label{eq:pn_post}
\end{equation}
After transmission over the wireless channel, the receiver observes \(\hat{\bar{\mathbf{z}}}_{\mathrm{tx}}\), i.e., the channel-corrupted version of \(\bar{\mathbf{z}}_{\mathrm{tx}}\) with noise or fading applied to the active channel-input symbols. The receiver then reconstructs a fixed-size complex latent \(\tilde{\mathbf{z}}\) by unpacking the received prefixes in token order and zero-filling all discarded suffix positions:
\begin{equation}
\tilde{\mathbf{z}}
=
\mathrm{unpack}\!\left(\hat{\bar{\mathbf{z}}}_{\mathrm{tx}},\boldsymbol{\tau}\right)
\in\mathbb{C}^{S\times (C/2)}.
\label{eq:z_tilde_unpack}
\end{equation}
The reconstructed image is then obtained as \(\hat{\mathbf{x}}=f_d(\tilde{\mathbf{z}};\boldsymbol{\phi})\).

\begin{figure}[t]
\centering
\includegraphics[width=0.84\linewidth]{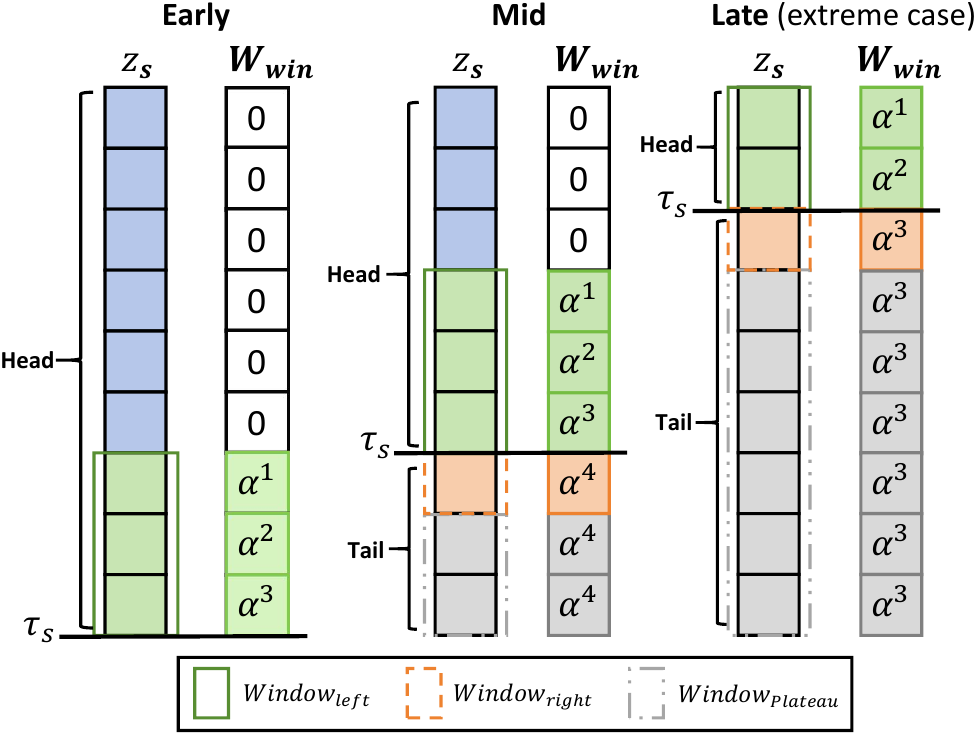}
\caption{\textbf{Sliding-window geometric L1-loss weights for tail-contiguous sparsity.} For each token \(s\), \(\mathbf{z}_s\) denotes its feature-channel vector, and \(\mathbf{W}_{\mathrm{win}}^{(s)}\) contains the L1-loss weights used in the sparsity regularizer around the current termination index \(\tau_s\). Early, Mid, and Late denote representative stages at the beginning, middle, and end of training, showing the progressive movement of \(\tau_s\) toward the head and the formation of a zero tail.}
\label{fig:l1_window}
\end{figure}

\subsection{Sliding-window weighted \texorpdfstring{L1}{L1} regularization}
\label{sec:win_l1_tail}

Tail-contiguous sparsity should reduce side information without unnecessarily weakening reconstruction. We therefore concentrate the L1 pressure near the tail of each active feature-channel prefix rather than over the whole latent vector. This design has two direct communication advantages. First, a contiguous zero tail yields a structured sparse pattern, so the threshold-derived active feature-channel count of each token can be encoded by one termination index instead of a dense mask. Second, because most head symbols remain unconstrained, the latent representation preserves the main reconstruction-relevant features. A global L1 penalty is not suitable for this purpose because it mainly produces unstructured shrinkage and also penalizes all candidate channel-input symbols, including informative head entries that should remain active. Instead, we use a sliding-window weighted L1 regularizer whose window is updated from the current threshold-derived \(\tau_s\), assigning zero penalty to the preserved head, geometrically increasing penalty around the current truncation boundary, and a saturated penalty to the remaining tail. Fig.~\ref{fig:l1_window} illustrates the corresponding L1-loss weight vector.

For each token \(s\), we define the sliding-window L1-loss weight vector \(\mathbf{W}_{\mathrm{win}}^{(s)}=[w_c^{(s)}]_{c=1}^{C/2}\),
where \(w_c^{(s)}\) is the loss weight assigned to the \(c\)-th complex symbol of the token feature-channel vector \(\mathbf{z}_s\). Let \(L_{\ell}\) and \(L_r\) denote the left and right window lengths, respectively. Here \(L_{\ell}\) controls how many symbols immediately before the current boundary receive growing sparsity pressure, whereas \(L_r\) controls the post-boundary transition length before the L1-loss weight reaches a saturated tail plateau. We define the window boundaries as
\begin{equation}
 \begin{aligned}
 c_0^{(s)}&=\max\{1,\tau_s-L_{\ell}+1\},\\
 c_1^{(s)}&=\min\left\{\frac{C}{2},\tau_s+L_r\right\}.
 \end{aligned}
\end{equation}
The resulting three-region loss-weight rule is as follows. Entries before \(c_0^{(s)}\) receive zero loss weight and therefore no sparsity penalty, which preserves the head symbols that are already important for transmission. Between \(c_0^{(s)}\) and \(c_1^{(s)}\), the weight increases geometrically; after \(c_1^{(s)}\), it remains at the saturated plateau value, so the tail is still suppressed rather than left unconstrained. This rule is written as
\begin{equation}
\begin{aligned}
w_c^{(s)}=
\begin{cases}
0, & c<c_0^{(s)},\\[3pt]
\alpha^{\,c-c_0^{(s)}+1}, & c_0^{(s)}\le c\le c_1^{(s)},\\[3pt]
\alpha^{\,c_1^{(s)}-c_0^{(s)}+1}, & c>c_1^{(s)},
\end{cases}
\qquad \alpha>1.
\end{aligned}
\end{equation}
where the geometric growth factor \(\alpha>1\) determines how aggressively the L1-loss weight increases from the preserved head toward the suppressed tail.
The resulting sliding-window weighted L1 loss is
\begin{equation}
\mathcal{L}_{\mathrm{win}\text{-}\mathrm{L1}}
=\lambda_{\mathrm{L1}}\sum_{s=1}^{S}\sum_{c=1}^{C/2} w_c^{(s)}\,|z_{s,c}|,
\label{eq:L1}
\end{equation}
where \(\lambda_{\mathrm{L1}}>0\) sets the global sparsity pressure associated with the selected average-CBR operating point. It does not prescribe token lengths directly: for a fixed \(\lambda_{\mathrm{L1}}\), the encoder output remains content dependent, and thresholding yields different \(\tau_s\) across tokens and images. During training, \(\tau_s\) is re-estimated from the current sparsity pattern of each token. As the weighted L1 term progressively suppresses tail entries, \(\tau_s\) shifts from the tail toward the head, forming an increasingly long contiguous zero tail, as illustrated by the Early--Mid--Late stages in Fig.~\ref{fig:l1_window}.

In fixed-rate training for TS-JSCC w/o SA\&RA, \(\lambda_{\mathrm{L1}}\) is constant across all samples and equals a run-level base coefficient \(\lambda_{\mathrm{base}}\), which sets the average-CBR scale requested by that model. In adaptive training for TS-JSCC (RA) and TS-JSCC (SA\&RA), a random multiplier \(\rho\) varies the rate condition by setting \(\lambda_{\mathrm{L1}}=\rho\,\lambda_{\mathrm{base}}\) for the sparsity loss and \(\lambda_{\mathrm{norm}}=\rho/\lambda_{\max}\) for the regulating module input. The concrete sampling strategy for \(\rho\) is provided in the Supplementary Material. The left window \(L_{\ell}\) specifies how many feature-channel symbols at the tail of the current active prefix receive nonzero sparsity regularization. A larger \(L_{\ell}\) exposes more near-tail symbols to the weighted L1 penalty and therefore accelerates sparsification, but it can also reduce reconstruction quality by constraining too many useful symbols. A smaller \(L_{\ell}\) weakens the structural pressure, which often slows down zero-tail formation and can make training less stable. The right window \(L_r\) specifies how many feature-channel symbols immediately after the current termination index still use geometrically increasing L1-loss weights; all later tail symbols share the saturated plateau weight and remain suppressed.

\subsection{Hard zeroing with magnitude-aware STE (mSTE)}
\label{sec:mste}
\label{sec:surrogate_grad}

As shown in Fig.~\ref{fig:overview_framework}, actual sparse transmission is realized by hard thresholding followed by termination-index-based tail truncation. During training, however, explicitly packing the active prefixes, truncating the tails, and unpacking them with receiver-side zero padding would introduce discrete operations that block gradient propagation. We therefore keep the latent tensor at its fixed shape and replace the tail positions that would not be selected for transmission with exact zeros. This zero-filled forward representation is the same fixed-shape tensor delivered to the decoder after active-prefix transmission and receiver-side zero filling, while avoiding non-differentiable packing and unpacking inside the training graph.

The zero-filled representation is implemented by a hard-threshold mask that keeps sufficiently strong entries and sets the remaining positions to zero. Given the complex latent \(\mathbf{z}\in\mathbb{C}^{S\times (C/2)}\), this mask is
\begin{equation}
m_{s,c}=\mathbb{I}\!\left(|z_{s,c}|\ge \epsilon\right),
\qquad
s\in\{1,\ldots,S\},\;
c\in\left\{1,\ldots,\frac{C}{2}\right\},
\label{eq:hz_mask}
\end{equation}
and hard zeroing is performed element-wise as
\begin{equation}
z^{\mathrm{hz}}_{s,c}=z_{s,c}\,m_{s,c},
\label{eq:hz_forward}
\end{equation}
where \(m_{s,c}=1\) indicates an active symbol and \(m_{s,c}=0\) drops the entry to an exact zero. The mask in Eq.~\eqref{eq:hz_mask} directly determines the termination index in Eq.~\eqref{eq:tau}: \(\tau_s\) is the last active position of token \(s\). During actual transmission, we send only the prefix up to \(\tau_s\), not the full binary mask.

Hard masking in Eq.~\eqref{eq:hz_forward} blocks reconstruction-loss gradients for dropped entries. A standard straight-through estimator (STE) keeps the hard-zero forward pass but backpropagates through each dropped entry with an approximate unit gradient. This can be seriously misleading near the threshold. If \(|z_{s,c}|\approx \epsilon^{-}\) (i.e., the symbol magnitude is just below the threshold), the forward pass removes the symbol completely, so its actual contribution after masking is zero; however, a vanilla STE still backpropagates as if the symbol had remained active, which encourages the optimizer to trust a forward path that does not exist. By contrast, when \(|z_{s,c}|\approx 0\), keeping or dropping the symbol leads to almost the same forward effect, so retaining part of the gradient is much less harmful. We therefore use the proposed magnitude-aware STE to suppress gradients for dropped symbols whose magnitudes are close to \(\epsilon\), while retaining gradients for dropped symbols near zero. This reduces misleading updates for near-threshold masked entries.

For the scalar mSTE construction, let \(m_{s,c}=\mathbb{I}(|z_{s,c}|\ge\epsilon)\). The detach-based construction is
\begin{equation}
{
\begin{aligned}
\psi_{s,c}
&=z_{s,c}-\frac{z_{s,c}|z_{s,c}|}{2\epsilon},\\
z^{\mathrm{hz}}_{s,c}
&=m_{s,c}z_{s,c}
+\bigl(1-m_{s,c}\bigr)\bigl(\psi_{s,c}-\mathrm{sg}[\psi_{s,c}]\bigr).
\end{aligned}
}
\label{eq:detach_trick}
\end{equation}
where \(\mathrm{sg}[\cdot]\) denotes stop-gradient. This keeps the forward pass identical to hard zeroing while yielding the following magnitude-dependent surrogate gradient for masked entries:
\begin{equation}
\begin{aligned}
\frac{\partial \mathcal{L}}{\partial z_{s,c}}
\approx
\begin{cases}
\dfrac{\partial \mathcal{L}}{\partial z^{\mathrm{hz}}_{s,c}},
& |z_{s,c}|\ge \epsilon,\\[8pt]
\left(1-\dfrac{|z_{s,c}|}{\epsilon}\right)
\dfrac{\partial \mathcal{L}}{\partial z^{\mathrm{hz}}_{s,c}},
& |z_{s,c}|<\epsilon,
\end{cases}
\end{aligned}
\label{eq:mste_backward}
\end{equation}
which suppresses gradients near the threshold and retains them near zero. In other words, mSTE is conservative for masked entries whose magnitudes are close to \(\epsilon\), where a small magnitude change would switch the keep/drop decision, and preserves gradients mainly when the dropped entry is already negligible.

\begin{figure*}[t]
\centering
\includegraphics[width=0.90\textwidth]{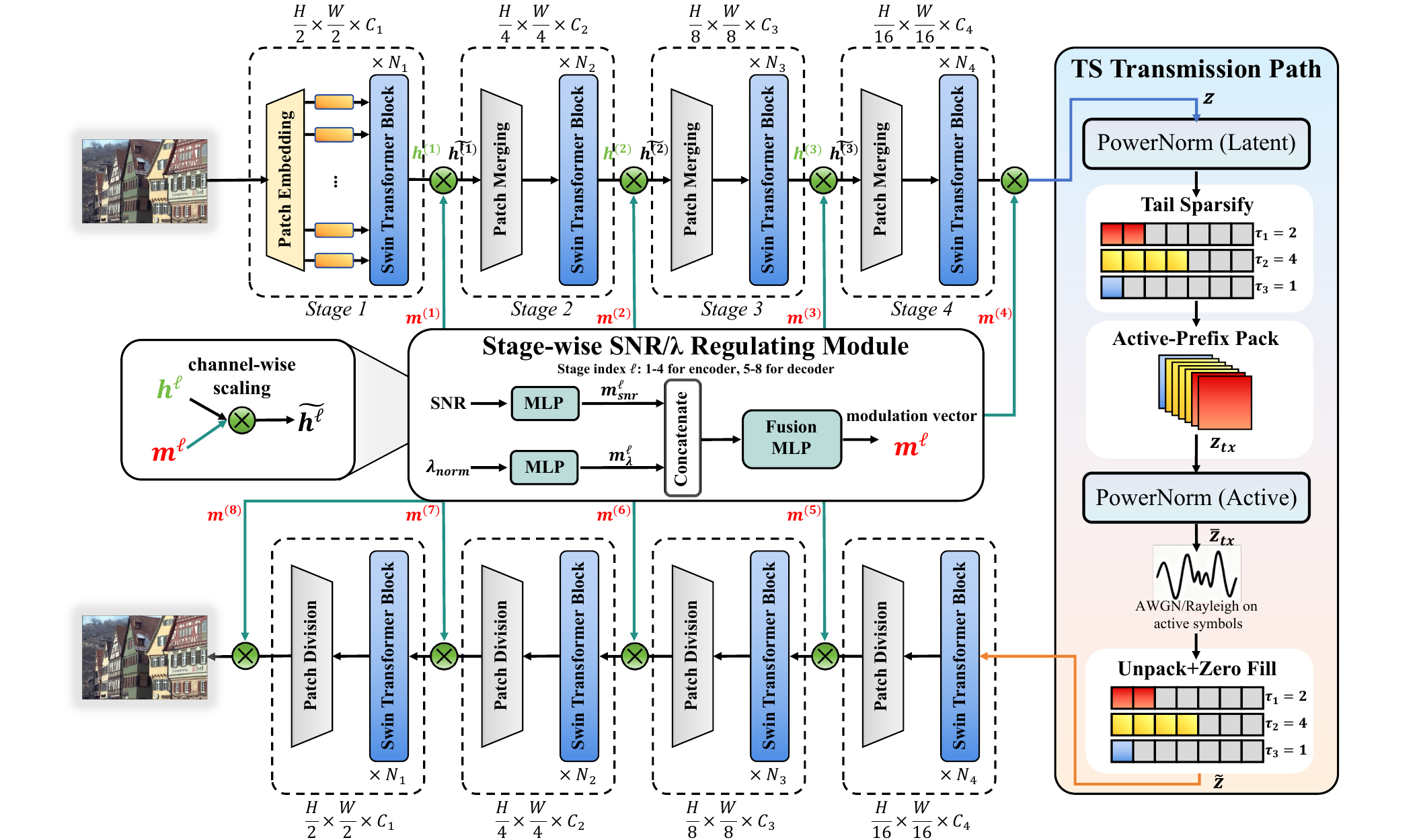}
\caption{Architecture of the full adaptive TS-JSCC (SA\&RA) model with stage-wise SNR/\(\lambda_{\mathrm{norm}}\) regulating modules and the tail-structured transmission path. The center block depicts one stage template; stage index \(\ell=1,\ldots,4\) corresponds to the encoder and \(\ell=5,\ldots,8\) corresponds to the decoder in data-flow order from the wireless channel output to the reconstructed image. Here \(C_i\) and \(N_i\) denote the stage feature-channel dimension and the number of Swin Transformer blocks, respectively.}
\label{fig:adaptive_modulation_architecture}
\end{figure*}

\subsection{Single-model multi-SNR and multi-rate adaptation}
\label{sec:single_model_adaptation}

Practical wireless image transmission rarely operates at a single fixed rate and a single fixed SNR. End devices, latency constraints, and link quality may all change over time, so a deployable learned communication system should cover a broad set of transmission rates and wireless channel conditions within a single model, implemented as one checkpoint at inference time, rather than retraining a dedicated checkpoint for every operating point. To meet this requirement, we condition both the encoder and the decoder on two inputs to the regulating modules: the SNR condition \(\gamma\) and the normalized sparsity coefficient \(\lambda_{\mathrm{norm}}\). In fixed-rate training for TS-JSCC w/o SA\&RA, the loss-side sparsity coefficient is simply \(\lambda_{\mathrm{L1}}=\lambda_{\mathrm{base}}\), where \(\lambda_{\mathrm{base}}\) is a run-level base coefficient that remains constant and sets the global average-CBR scale. In adaptive training for TS-JSCC (RA) and TS-JSCC (SA\&RA), a randomly sampled multiplier \(\rho\) is introduced so that the same checkpoint learns to respond to different \(\lambda_{\mathrm{norm}}\) values and thereby covers multiple average-CBR levels. It also forms the loss-side sparsity coefficient \(\lambda_{\mathrm{L1}}=\rho\,\lambda_{\mathrm{base}}\). The normalized rate-control input is computed as
\begin{equation}
\lambda_{\mathrm{norm}}=\frac{\rho}{\lambda_{\max}}.
\end{equation}
The value \(\lambda_{\mathrm{norm}}\) is then fed to the regulating modules as the rate-control input, where \(\lambda_{\max}\) is the maximum multiplier used for normalization. Larger \(\lambda_{\mathrm{L1}}\) values impose stronger tail suppression and therefore lower average \cbr{}, whereas smaller \(\lambda_{\mathrm{L1}}\) values preserve longer active prefixes for higher-fidelity reconstruction. The sampling scheme used to draw \(\rho\) is described in the Supplementary Material. In the fixed-rate reference models, the operating point is fixed by \(\lambda_{\mathrm{base}}\) alone and no adaptive \(\lambda_{\mathrm{norm}}\) input is required.

Fig.~\ref{fig:adaptive_modulation_architecture} illustrates the full adaptive TS-JSCC (SA\&RA) architecture with stage-wise regulating modules for joint SNR- and rate-adaptive operation. Let \(\mathbf{h}^{(\ell)}\in\mathbb{R}^{S_{\ell}\times C_{\ell}}\) denote the feature tensor at stage \(\ell\), where \(S_{\ell}\) is the stage-dependent token count, \(C_{\ell}\) is the embedding dimension, \(\ell=1,\ldots,4\) correspond to the encoder stages, and \(\ell=5,\ldots,8\) correspond to the decoder stages. The SNR condition is denoted by \(\gamma\) and labeled as SNR in Fig.~\ref{fig:adaptive_modulation_architecture}. Because \(\gamma\) and \(\lambda_{\mathrm{norm}}\) are sample-level operating conditions, each regulating module generates one modulation vector per image sample and broadcasts it to all tokens at the corresponding stage. It therefore sets the global rate/SNR operating point, whereas token-wise allocation is still determined by the TS module through threshold-derived active prefixes. A dedicated SNR branch, a dedicated \(\lambda_{\mathrm{norm}}\) branch, and a fusion branch---implemented as separate lightweight MLPs with the same hidden-layer template but branch-specific input dimensions and parameters---combine the two conditioning variables into a shared modulation vector:
\begin{align}
\mathbf{m}^{(\ell)}_{\mathrm{snr}} &= f^{(\ell)}_{\mathrm{snr}}(\gamma),\\
\mathbf{m}^{(\ell)}_{\lambda} &= f^{(\ell)}_{\lambda}(\lambda_{\mathrm{norm}}),\\
\mathbf{m}^{(\ell)} &= f^{(\ell)}_{\mathrm{fuse}}\!\left([\mathbf{m}^{(\ell)}_{\mathrm{snr}};\mathbf{m}^{(\ell)}_{\lambda}]\right),
\end{align}
which is then broadcast to all tokens and applied through feature-wise (embedding-dimension) scaling:
\begin{equation}
\tilde{\mathbf{h}}^{(\ell)}_s=\mathbf{h}^{(\ell)}_s\odot \mathbf{m}^{(\ell)},\qquad s=1,\ldots,S_{\ell}.
\end{equation}
Each encoder or decoder stage uses its own SNR MLP, \(\lambda_{\mathrm{norm}}\) MLP, and fusion MLP; these MLPs are not shared across stages. The output length of each regulating module is set to the feature dimension \(C_{\ell}\) of that stage, so \(\mathbf{m}^{(\ell)}\) can directly scale \(\mathbf{h}^{(\ell)}_s\) along the embedding dimension. Compared with SwinJSCC w/ RA and SwinJSCC w/ SA\&RA, the proposed design uses only lightweight per-stage regulating modules and feature-wise scaling, with marginal additional complexity as quantified in Table~\ref{tab:complexity_summary}, to cover wide CBR and SNR ranges. The regulating modules and the TS module work jointly: the regulating modules set the sample-level rate/SNR condition, while the TS module performs token-wise content-adaptive feature-channel allocation.

\subsection{Overall training objective}

The final training loss simply adds the reconstruction term to the sliding-window weighted sparsity term:
\begin{equation}
\mathcal{L}
=
\mathcal{L}_{\mathrm{rec}}
+
\mathcal{L}_{\mathrm{win}\text{-}\mathrm{L1}},
\end{equation}
where \(\mathcal{L}_{\mathrm{rec}}\) denotes the reconstruction term associated with the distortion objective in Eq.~\eqref{eq:deepjscc_obj}, and \(\mathcal{L}_{\mathrm{win}\text{-}\mathrm{L1}}\) is the sliding-window weighted sparsity term in Eq.~\eqref{eq:L1}. The choices of \(\lambda_{\mathrm{L1}}\), \(\lambda_{\mathrm{base}}\), \(\rho\), and \(\lambda_{\mathrm{norm}}\) for the fixed-rate and adaptive models are specified in Secs.~\ref{sec:single_model_adaptation} and~\ref{sec:exp_protocol}.

\section{Experimental Setup}
\label{sec:experimental_setup}
\label{sec:exp_protocol}

We evaluate the proposed framework on both low- and high-resolution image transmission tasks. For low resolution, we use CIFAR-10 \cite{CIFAR10} for both training and evaluation. For high resolution, we train on DIV2K \cite{DIV2K} with random \(256\times256\) crops and test on Kodak \cite{Kodak} and CLIC2021 \cite{CLIC2021}; at test time, each image is center cropped to the nearest size divisible by \(128\). Both AWGN and Rayleigh fading channels are evaluated.

The detailed experimental settings are provided in Supplementary Table~\ref{tab:setup_summary}. Following the naming convention of SwinJSCC, we report three TS-JSCC variants. TS-JSCC w/o SA\&RA uses the same Swin Transformer encoder--decoder active path as SwinJSCC w/o SA\&RA and does not include SNR- or rate-aware regulating modules; the difference is the proposed tail-structured L1 regularization and active-prefix transmission rule. TS-JSCC (RA) adds lightweight stage-wise regulating modules driven by \(\lambda_{\mathrm{norm}}\), and is trained at a fixed SNR while sampling different sparsity/rate conditions so that one model covers multiple average CBRs. TS-JSCC (SA\&RA) further feeds both the SNR condition and \(\lambda_{\mathrm{norm}}\) into the regulating modules, yielding a fully adaptive model across channel-quality and rate variations. For SA\&RA training, SNRs are sampled uniformly from \([0,13]\) dB under both AWGN and Rayleigh fading; the rate-adaptive protocol is provided in the Supplementary Material. NTSCC++ online uses 20 Adam updates per image, whereas all other learned-JSCC curves use feed-forward inference.

For separation-based baselines, BPG or HPCM bitstreams are transmitted with SNR-specific LDPC+QAM modes. The AWGN/Rayleigh schedules are listed in Supplementary Table~\ref{tab:ldpc_qam_schedule}; Fig.~\ref{fig:mcs_cliff} uses the AWGN modes. For AWGN, QAM soft demodulation uses the corresponding additive-noise likelihood. For Rayleigh fading, we follow the same amplitude-fading model as the learned JSCC evaluations, \(y=hx+n\) with real nonnegative \(h\) and \(\mathbb{E}[h^2]=1\). Since instantaneous fading coefficients are not provided to the separated receiver, its QAM demapper uses the known Rayleigh fading distribution to compute statistical soft likelihoods before LDPC decoding.

\section{Results and Discussion}
\label{sec:results_discussion}

\subsection{PSNR--CBR performance under \texorpdfstring{\textbf{AWGN}}{AWGN} channels}

\begin{figure*}[t]
\centering
\subfloat[Kodak.]{
\includegraphics[width=0.31\textwidth]{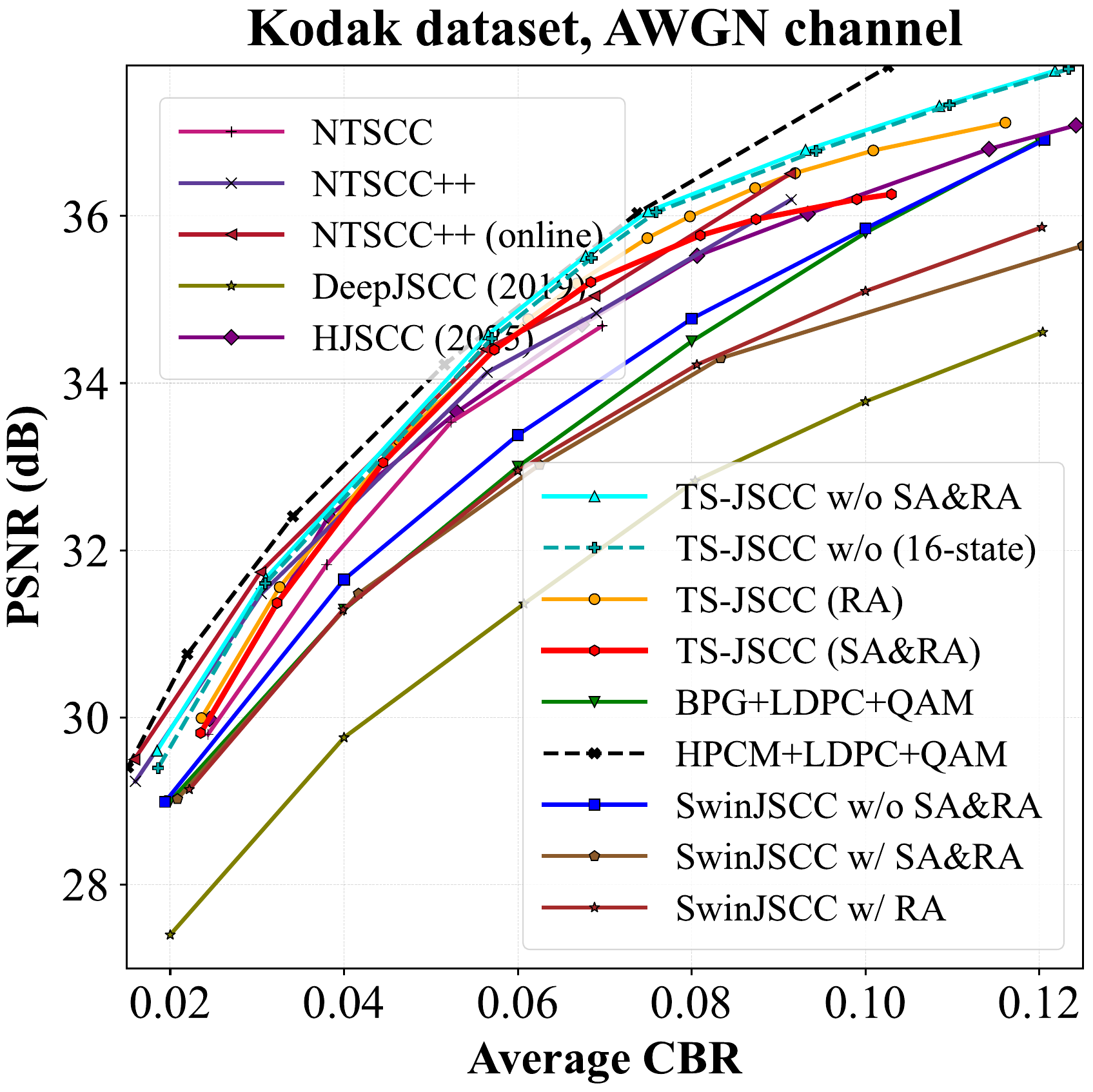}
\label{fig:kodak_awgn_cbr}
}
\hfil
\subfloat[CLIC2021.]{
\includegraphics[width=0.31\textwidth]{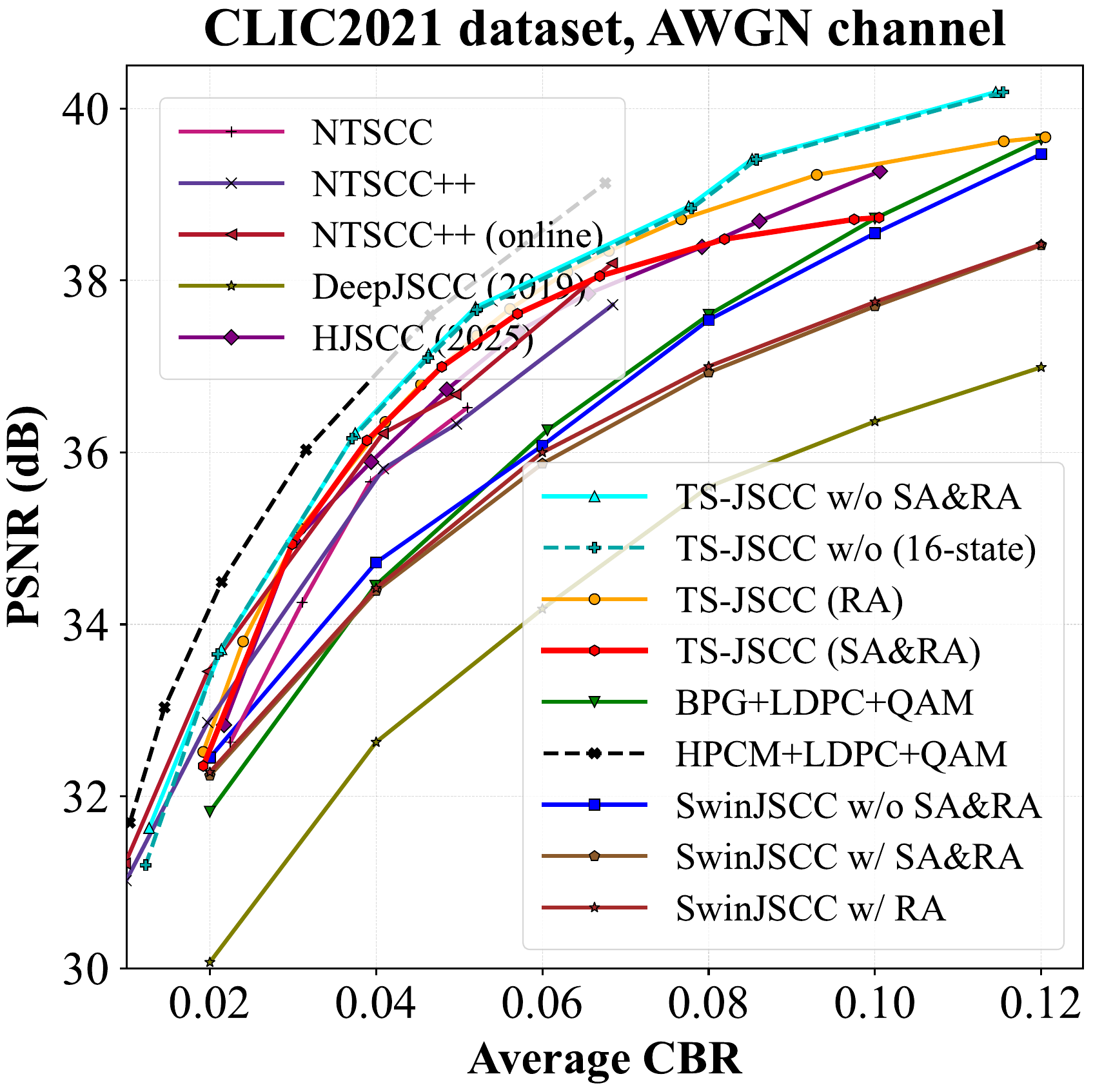}
\label{fig:clic_awgn_cbr}
}
\hfil
\subfloat[CIFAR-10.]{
\includegraphics[width=0.31\textwidth]{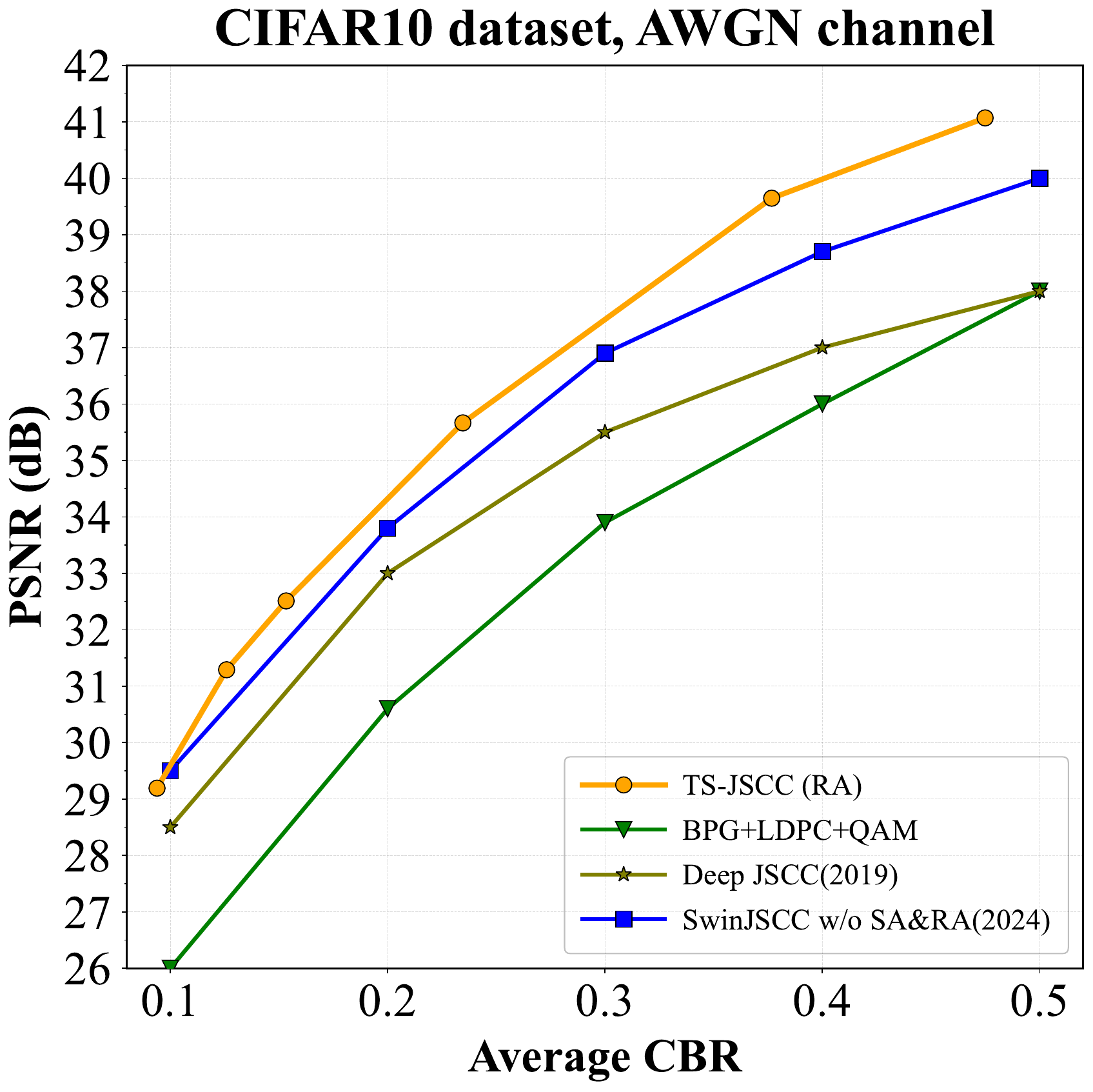}
\label{fig:cifar_awgn_cbr}
}
\caption{AWGN PSNR--CBR performance at fixed \(\snr=10\) dB. RA denotes a single-model rate-adaptive setting covering multiple average CBRs, and SA\&RA denotes a single-model setting jointly adaptive to SNR and rate.}
\label{fig:awgn_foundation}
\end{figure*}

\begin{figure}[t]
\centering
\subfloat[Kodak.]{
\includegraphics[width=0.46\columnwidth]{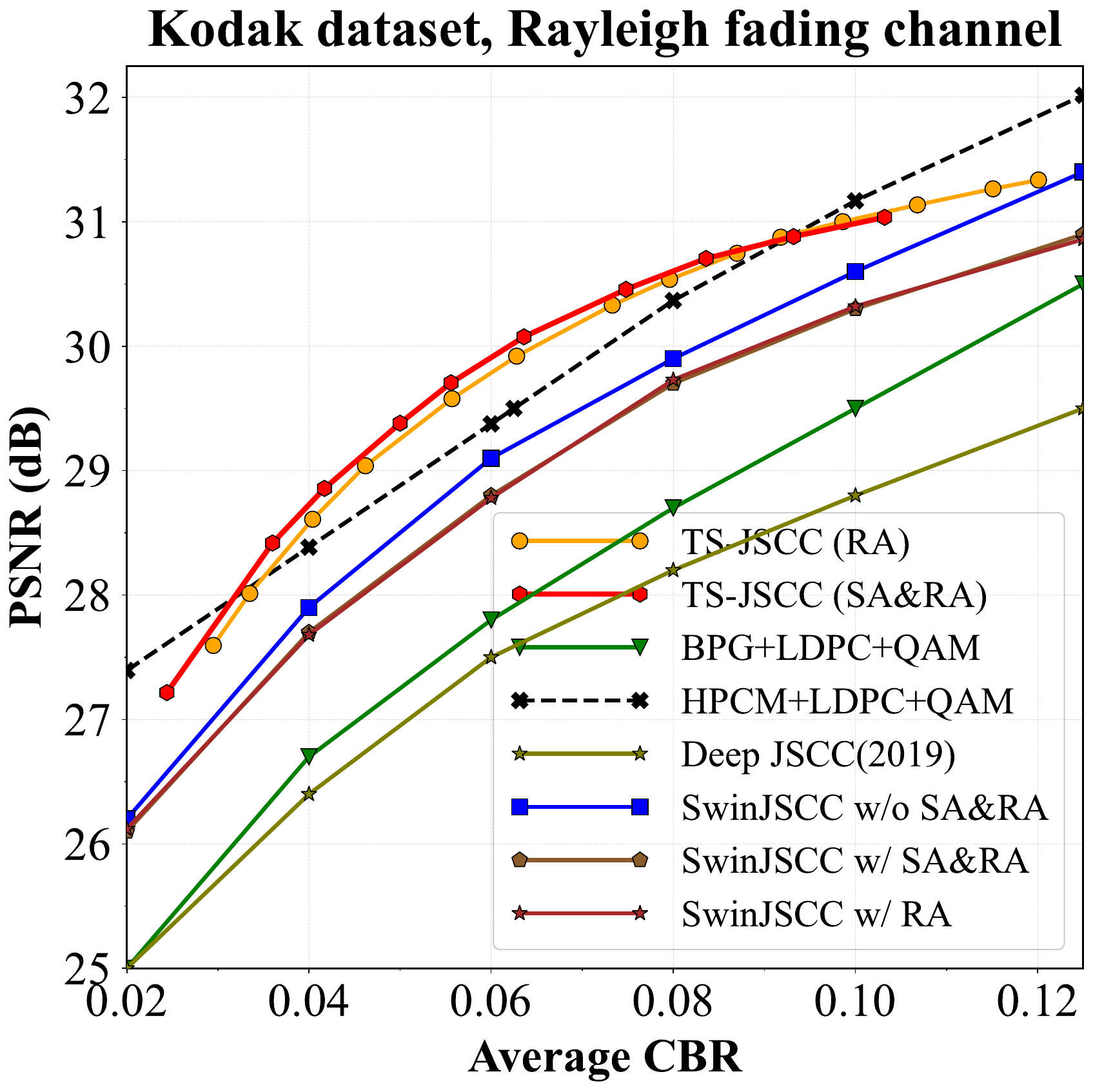}
\label{fig:kodak_rayleigh_cbr}
}%
\hfill
\subfloat[CLIC2021.]{
\includegraphics[width=0.46\columnwidth]{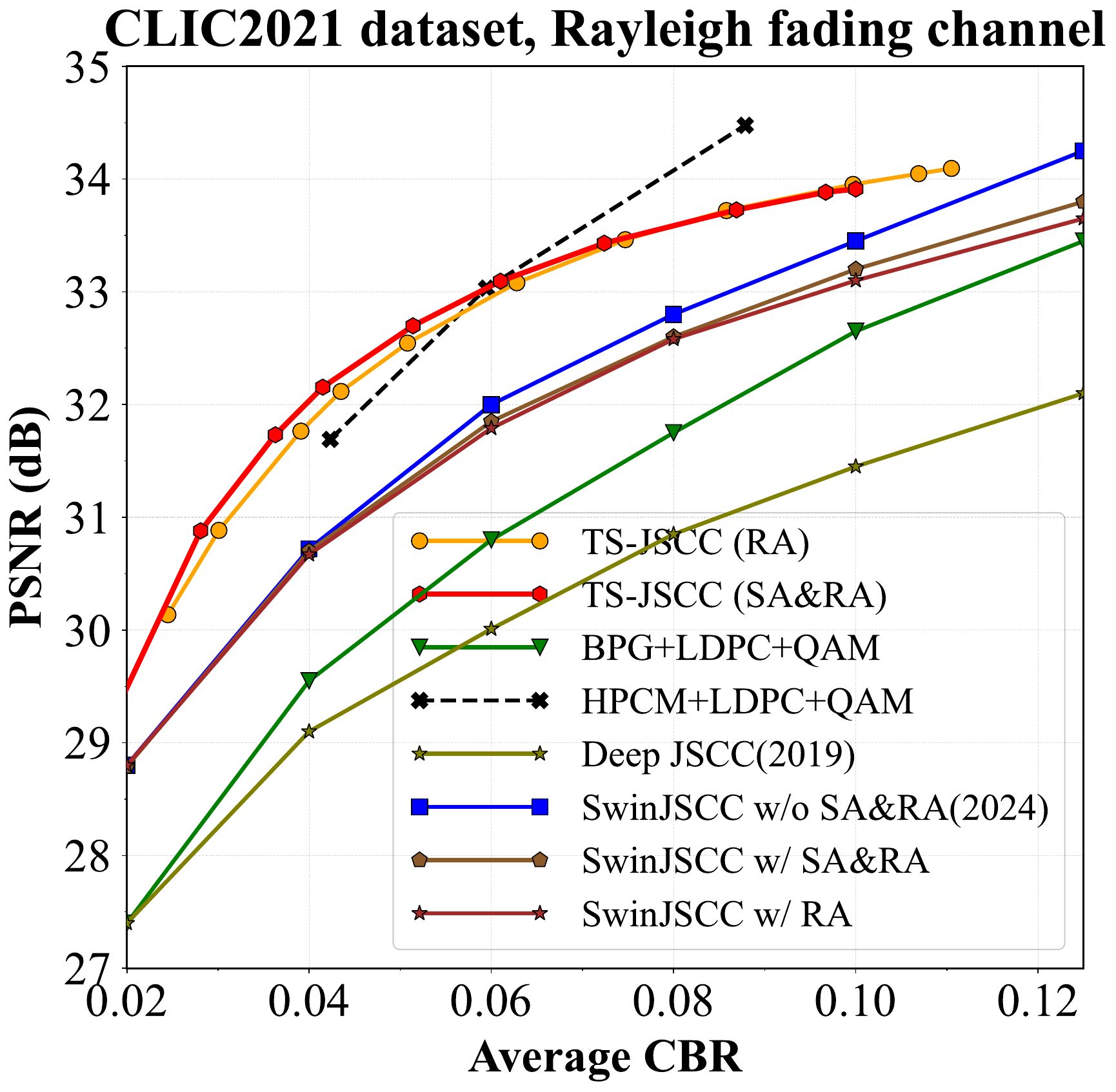}
\label{fig:clic_rayleigh_cbr}
}
\caption{Rayleigh PSNR--CBR performance at fixed \(\snr=3\) dB.}
\label{fig:rayleigh_cbr}
\end{figure}

\begin{figure}[t]
\centering
\subfloat[AWGN.]{
\includegraphics[width=0.46\columnwidth]{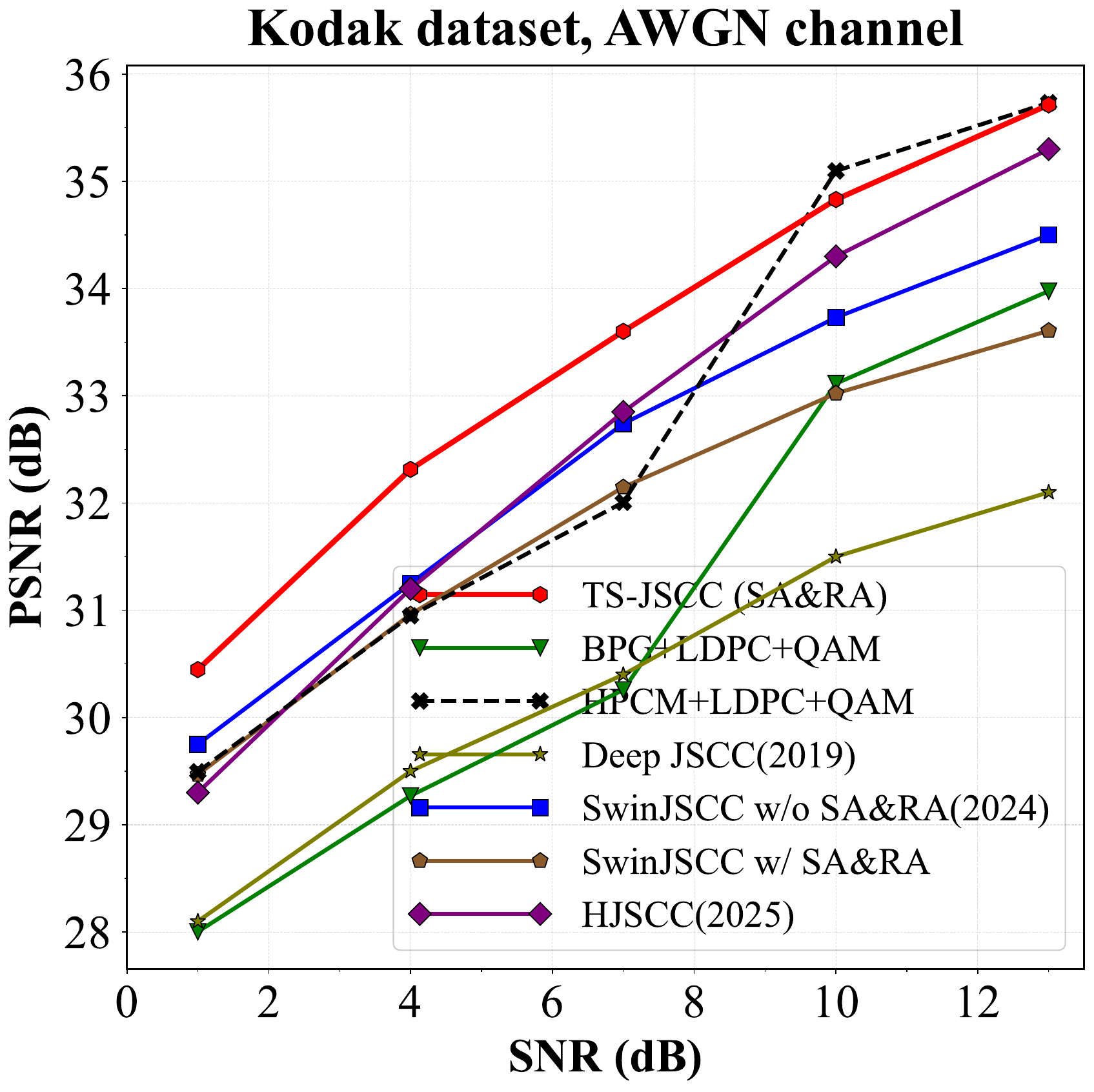}
\label{fig:kodak_awgn_snr}
}%
\hfill
\subfloat[Rayleigh fading.]{
\includegraphics[width=0.46\columnwidth]{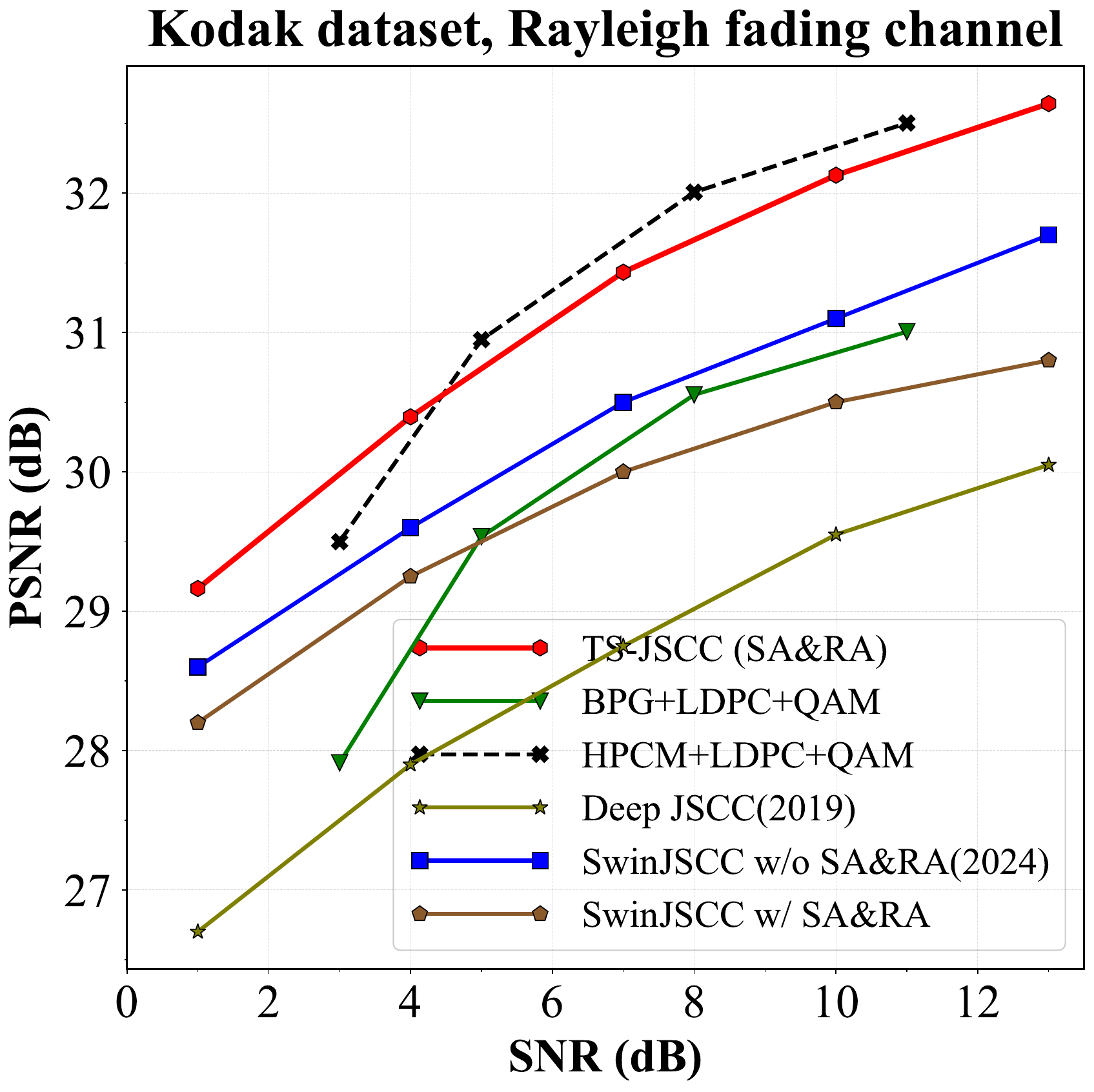}
\label{fig:kodak_rayleigh_snr}
}
\caption{Kodak PSNR--SNR performance at fixed average \(\cbr \approx 0.0625\).}
\label{fig:snr_robustness}
\end{figure}

\begin{figure}[t]
\centering
\subfloat[CBR sweep.]{
\includegraphics[width=0.46\columnwidth]{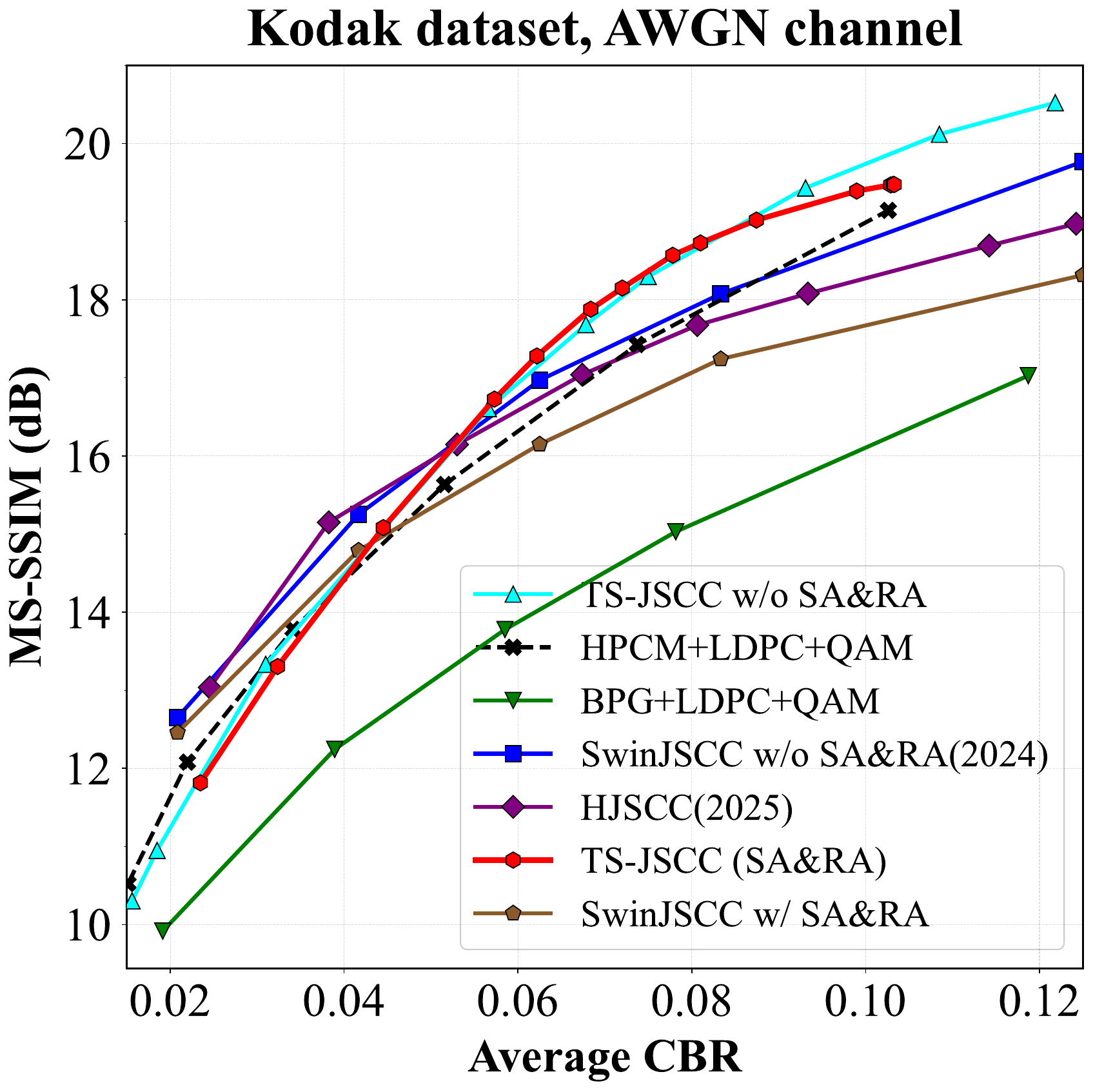}
\label{fig:kodak_awgn_cbr_msssim}
}%
\hfill
\subfloat[SNR sweep.]{
\includegraphics[width=0.46\columnwidth]{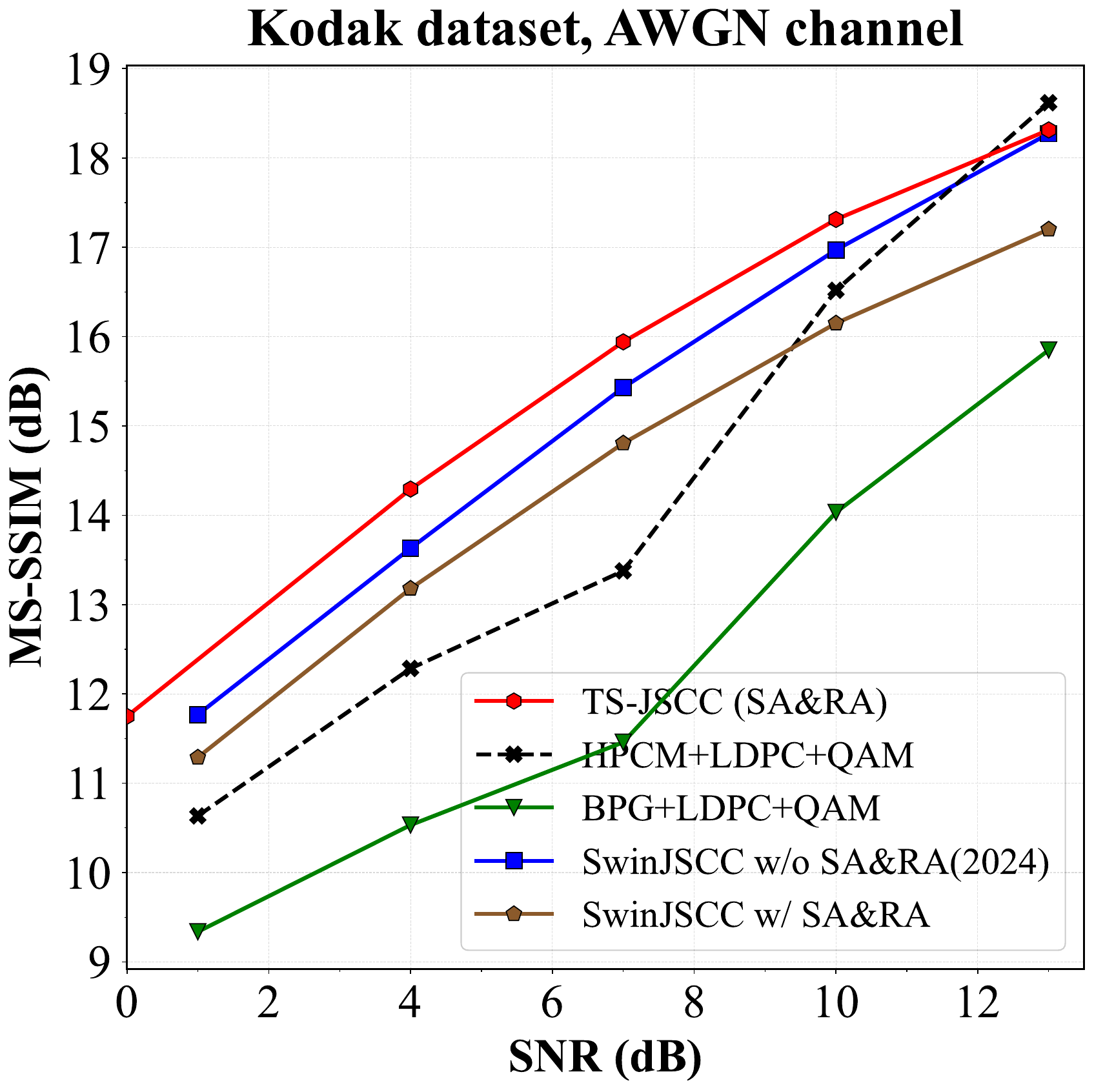}
\label{fig:kodak_awgn_snr_msssim}
}
\caption{Kodak MS-SSIM (dB) performance under AWGN. The CBR sweep compares fixed-rate and adaptive schemes at \(\snr=10\) dB, while the SNR sweep compares fixed-CBR operation at average \(\cbr \approx 0.0625\). All curves use MSE-trained models.}
\label{fig:msssim_awgn}
\end{figure}

\begin{figure}[t]
\centering
\includegraphics[width=0.90\linewidth]{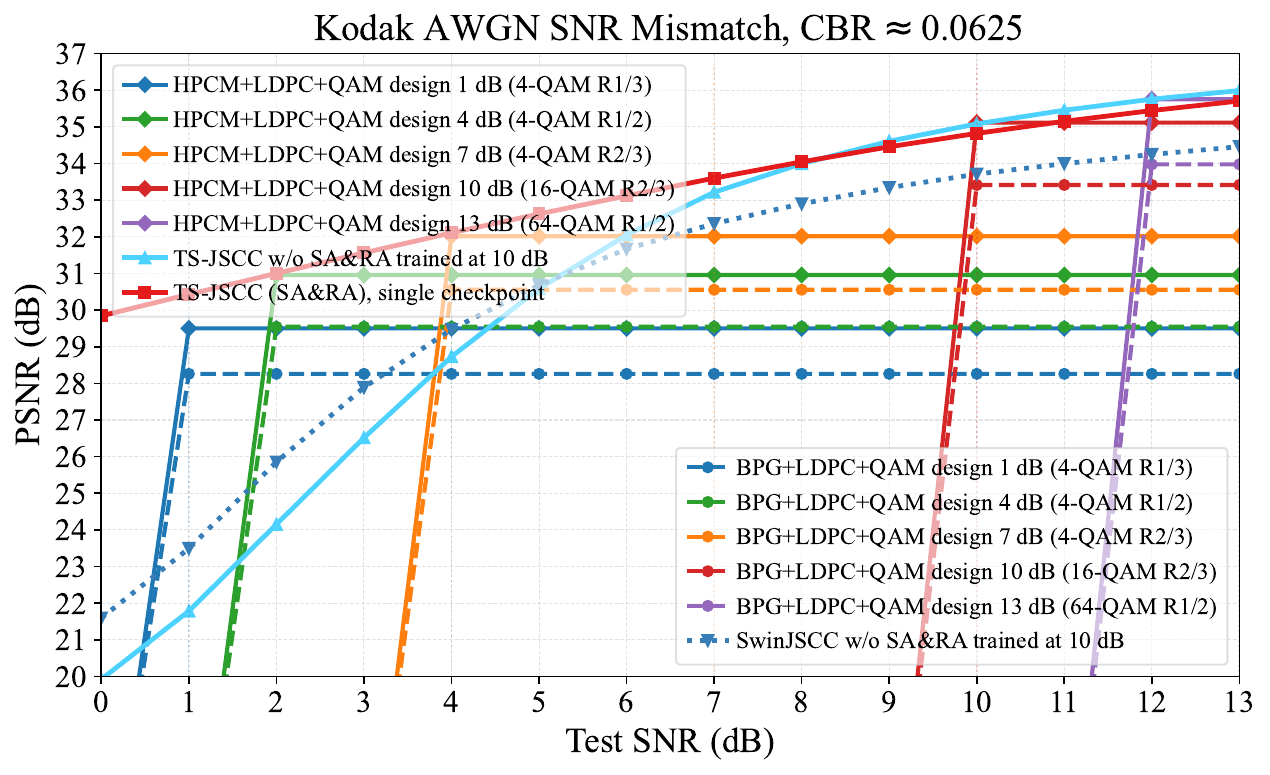}
\caption{SNR-mismatch behavior on Kodak at average \(\cbr \approx 0.0625\). BPG+LDPC+QAM and HPCM+LDPC+QAM are separated digital baselines, TS-JSCC w/o SA\&RA and SwinJSCC w/o SA\&RA use \(10\) dB-trained fixed-SNR models, and TS-JSCC (SA\&RA) uses a single SNR-adaptive model; digital separated curves exhibit quality cliffs below their design SNRs.}
\label{fig:mcs_cliff}
\end{figure}

Fig.~\ref{fig:awgn_foundation} compares PSNR--CBR trade-offs under AWGN at \(\snr=10\) dB. In Figs.~\ref{fig:kodak_awgn_cbr} and~\ref{fig:clic_awgn_cbr}, TS-JSCC w/o SA\&RA achieves the highest PSNR among the compared learned-JSCC methods over most of the evaluated CBR range. NTSCC++ (online) has a slight advantage only at the lowest CBRs, but requires 20 image-specific Adam updates before inference, incurring additional latency and computational cost. HPCM+LDPC+QAM is slightly higher at several AWGN points, but this separated curve is an optimistic matched-mode reference, and the reported PSNR is valid only when the entropy-coded HPCM/BPG bitstream is recovered successfully. These MCS choices are reliability-oriented rather than error-free guarantees. In practical transmission, residual LDPC/QAM errors can corrupt the bitstream, so the actual visual quality is expected to be lower than the plotted curve and may collapse abruptly instead of degrading smoothly. This sensitivity is illustrated by the SNR-mismatch cliff behavior in Fig.~\ref{fig:mcs_cliff}. By contrast, TS-JSCC provides comparable AWGN reconstruction quality with a one-shot learned JSCC encoder--decoder, lower model complexity as quantified in Table~\ref{tab:complexity_summary}, and smoother degradation when the test channel deviates from the design SNR.

The gains over SwinJSCC w/o SA\&RA further clarify the role of dynamic resource allocation. Both HJSCC and TS-JSCC introduce content-adaptive symbol usage on top of the SwinJSCC w/o SA\&RA reference backbone, and both improve clearly over the uniform-allocation baseline. For TS-JSCC, this improvement is obtained without an entropy model, rate map, or auxiliary allocation network: the L1-driven tail sparsification converts uniform feature-channel usage into image-dependent active prefixes.

The adaptive curves reveal the expected specialization--flexibility trade-off. TS-JSCC (RA) and TS-JSCC (SA\&RA) lie slightly below TS-JSCC w/o SA\&RA, mirroring the trend of the SwinJSCC adaptive variants. The adaptation gap is small for TS-JSCC: TS-JSCC (RA) still outperforms HJSCC and SwinJSCC w/o SA\&RA over most evaluated CBRs, while TS-JSCC (SA\&RA) remains comparable to HJSCC and substantially above SwinJSCC w/ RA and SwinJSCC w/ SA\&RA. This behavior indicates that the proposed tail-structured sparsification improves resource utilization, and the proposed stage-wise lightweight SNR/rate regulating structure preserves much of this gain when one model is required to operate across multiple wireless channel conditions and rates. The low-resolution check in Fig.~\ref{fig:cifar_awgn_cbr} shows the same basic trend, with TS-JSCC remaining above SwinJSCC w/o SA\&RA on CIFAR-10 over the evaluated CBR range.

\subsection{PSNR--CBR performance under \texorpdfstring{\textbf{Rayleigh}}{Rayleigh} channels}

Fig.~\ref{fig:rayleigh_cbr} reports PSNR--CBR curves under Rayleigh fading at \(\snr=3\) dB, where each latent symbol is subject to random multiplicative attenuation before additive noise. TS-JSCC (RA) and TS-JSCC (SA\&RA) remain above the SwinJSCC adaptive baselines as well as SwinJSCC w/o SA\&RA across the evaluated CBR range, following the trend observed under AWGN. This result indicates that the proposed rate-allocation and lightweight adaptation design generalizes beyond additive-noise channels and retains its effectiveness under fading.

\subsection{PSNR--SNR robustness at fixed CBR}

Fig.~\ref{fig:snr_robustness} evaluates PSNR as a function of SNR at fixed average \(\cbr \approx 0.0625\). The proposed TS-JSCC (SA\&RA) uses a single model over the whole SNR range and achieves the highest PSNR among the compared learned-JSCC methods under both AWGN and Rayleigh fading, consistently surpassing the SwinJSCC adaptive variants and, more notably, even outperforming SwinJSCC w/o SA\&RA, whose points correspond to fixed-SNR operating models. BPG/HPCM+LDPC+QAM and SwinJSCC w/o SA\&RA use SNR-specific operating points, whereas TS-JSCC (SA\&RA) covers all tested SNRs with one adaptive model. Under AWGN, TS-JSCC (SA\&RA) exceeds HPCM+LDPC+QAM over a broad SNR range; under Rayleigh fading, it closely approaches the HPCM matched-mode points without SNR-specific model switching. As Table~\ref{tab:complexity_summary} shows, this continuous SNR adaptability is achieved with only marginal parameter and computation overhead relative to the TS-JSCC w/o SA\&RA backbone.

Fig.~\ref{fig:mcs_cliff} further illustrates the operating-point sensitivity of separated digital pipelines: when the realized SNR drops below the value for which the MCS was selected, BPG+LDPC+QAM and HPCM+LDPC+QAM can exhibit a pronounced cliff effect, whereas learned JSCC schemes degrade more smoothly. Thus, the separated points in Fig.~\ref{fig:snr_robustness} should be read as successful-recovery references rather than guaranteed error-free end-to-end performance. TS-JSCC (SA\&RA) uses the same model over all tested SNRs and maintains a smooth high-PSNR trajectory, showing the benefit of the proposed SNR-aware regulating structure under SNR mismatch.

\subsection{MS-SSIM evaluation without perceptual-loss training}
\label{sec:msssim_eval}

All learned JSCC curves in this subsection are obtained from MSE-trained models; MS-SSIM is used only as an additional evaluation metric, not as a training objective. Fig.~\ref{fig:msssim_awgn}(a) reports the CBR sweep at \(\snr=10\) dB. At low CBRs, all methods operate with very limited channel resources, so the full-image structural quality is constrained and the MS-SSIM margins are modest. In this regime, the content-adaptive allocation of TS-JSCC tends to preserve foreground and detail-rich regions while smoothing background regions more strongly. Since MS-SSIM averages luminance, contrast, and structural statistics over the whole image and across scales, background degradation can reduce the image-level score even when the ROI details remain visually stronger, as shown in Fig.~\ref{fig:visual_low}. As CBR increases, TS-JSCC has sufficient resources to recover both foreground details and background structures; its MS-SSIM advantage therefore becomes clear in the medium-to-high CBR regime and follows the PSNR trend. Fig.~\ref{fig:msssim_awgn}(b) further shows that, at \(\cbr \approx 0.0625\), TS-JSCC (SA\&RA) achieves the best learned MS-SSIM performance over the evaluated SNR range using a single model.

\begin{figure*}[!t]
\centering
\includegraphics[width=0.80\textwidth]{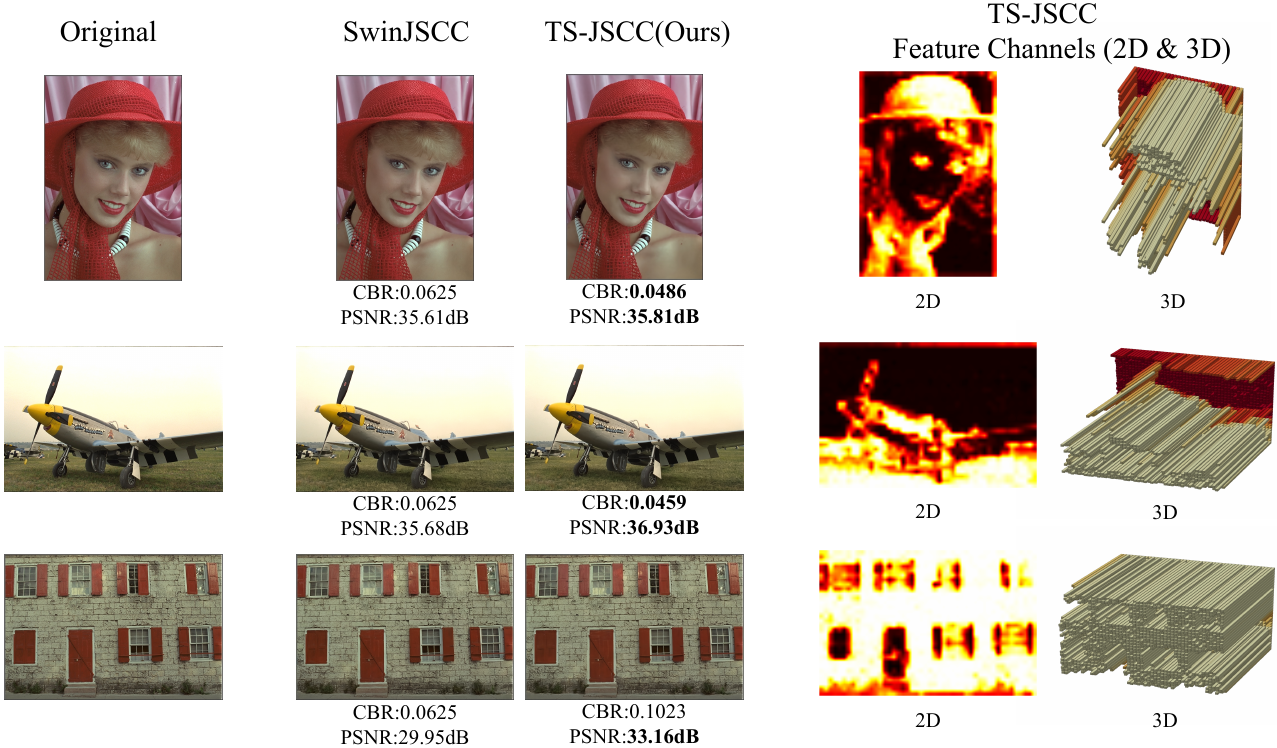}
\caption{\textbf{Automatic resource allocation visualization.} SwinJSCC w/o SA\&RA versus TS-JSCC w/o SA\&RA at \(\snr=10\) dB with matched average \(\cbr \approx 0.0625\). The numbers below each example are the image-specific CBR and PSNR. For TS-JSCC, the 2D/3D maps visualize token-wise feature-channel allocation, where brighter/taller regions indicate more active feature channels.}
\label{fig:visual_fixcbr}
\end{figure*}

\subsection{Qualitative allocation and ROI analysis}

\begin{figure*}[!t]
\centering
\includegraphics[width=0.82\textwidth,height=0.40\textheight,keepaspectratio]{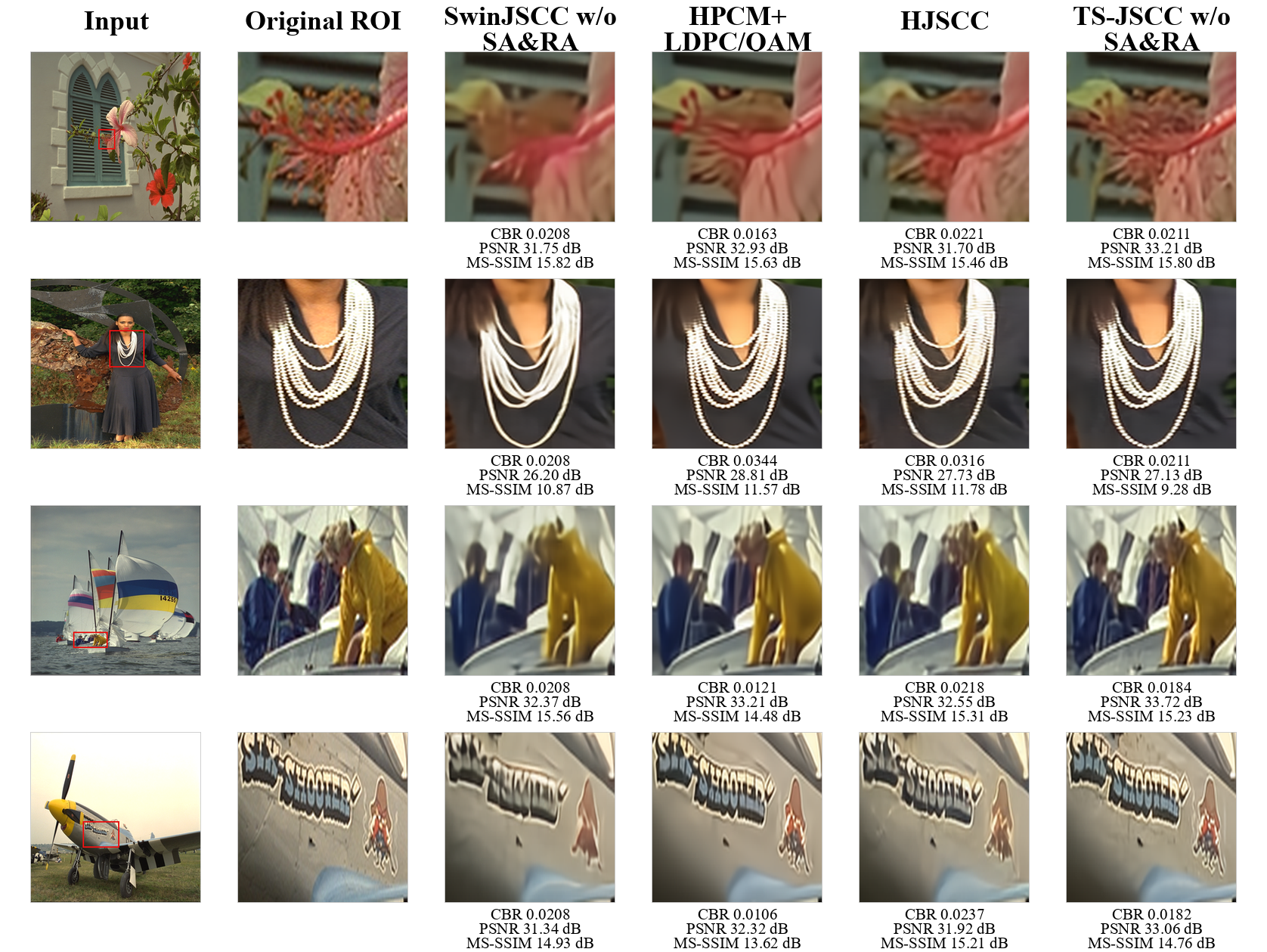}
\caption{\textbf{Low-bandwidth visual comparison} at \(\snr=10\) dB. Multiple transmission methods are compared, and zoomed regions of interest (ROIs) highlight local reconstruction details. The annotated CBR, PSNR, and MS-SSIM~(dB) values are full-image scores rather than ROI-only measurements.}
\label{fig:visual_low}
\end{figure*}

The qualitative evidence in Fig.~\ref{fig:allocation_compare_intro}, Fig.~\ref{fig:visual_fixcbr}, and Fig.~\ref{fig:visual_low} complements the quantitative curves. In Fig.~\ref{fig:visual_fixcbr}, both TS-JSCC and SwinJSCC use the w/o SA\&RA setting and share the same SwinJSCC backbone, so the comparison isolates the proposed allocation mechanism rather than network capacity. TS-JSCC performs automatic and content-adaptive allocation under the same average bandwidth budget: visually simpler images use lower image-wise CBR, whereas detail-rich images receive more resources. For the first two examples in Fig.~\ref{fig:visual_fixcbr}, TS-JSCC obtains higher PSNR with lower image-wise CBR than SwinJSCC; for the more detailed third example, it activates more feature-channel symbols and substantially improves reconstruction quality. This behavior is consistent with the bandwidth-utilization gains observed in the PSNR--CBR curves.

Fig.~\ref{fig:visual_low} focuses on detail recovery in the low-CBR, high-compression regime and compares learned JSCC with separated transmission baselines. SwinJSCC w/o SA\&RA lacks content-adaptive symbol allocation and therefore shows weaker ROI recovery in this low-bandwidth setting. Compared with HJSCC, TS-JSCC preserves finer local structures in the shown examples, including flower stamens, clear inter-pearl boundaries, facial detail, and color/edge fidelity in the aircraft sample. HPCM+LDPC+QAM can provide comparable local visual quality when its compressed bitstream is correctly decoded under the selected LDPC+QAM mode; however, under SNR or channel mismatch, the separated receiver may fail abruptly because of the cliff effect shown in Fig.~\ref{fig:mcs_cliff}, rather than degrading gradually as JSCC schemes usually do.

\subsection{Side-information overhead and model complexity}

\input{tables/tab03_sideinfo_overhead.tex}

\input{tables/tab04_complexity.tex}

Table~\ref{tab:sideinfo_overhead_highres} compares the additional CBR required to transmit side information without error on the high-resolution setting at \(\snr=10\) dB. For Global-\(\mathrm{L1}\), sparsification emerges at randomly distributed symbol locations, so the receiver must be informed of the full symbol-wise support via a dense mask, resulting in a large overhead. HJSCC also uses termination indices rather than a dense mask, but its multi-level architecture requires multiple \(\boldsymbol{\tau}\) signals and therefore accumulates a non-negligible overhead. In contrast, TS-JSCC needs only one compact termination index per token. Using a discrete side-information quantization principle as in the NTSCC family, the 16-state TS-JSCC option further reduces the capacity-equivalent overhead to \(0.00151\) CBR.

Table~\ref{tab:complexity_summary} reports model complexity under the common high-resolution setting. TS-JSCC w/o SA\&RA has the lowest parameter count (18.36M) and forward-path FLOPs (69.34G) among the compared learned-JSCC methods, while sharing the same backbone complexity as SwinJSCC w/o SA\&RA. Its lightweight regulating modules add only marginal storage and computational cost, so TS-JSCC (SA\&RA) retains strong deployment and inference efficiency while jointly adapting to rate and SNR.

\subsection{Ablation results}

\begin{figure}[t]
\centering
\includegraphics[width=0.68\linewidth]{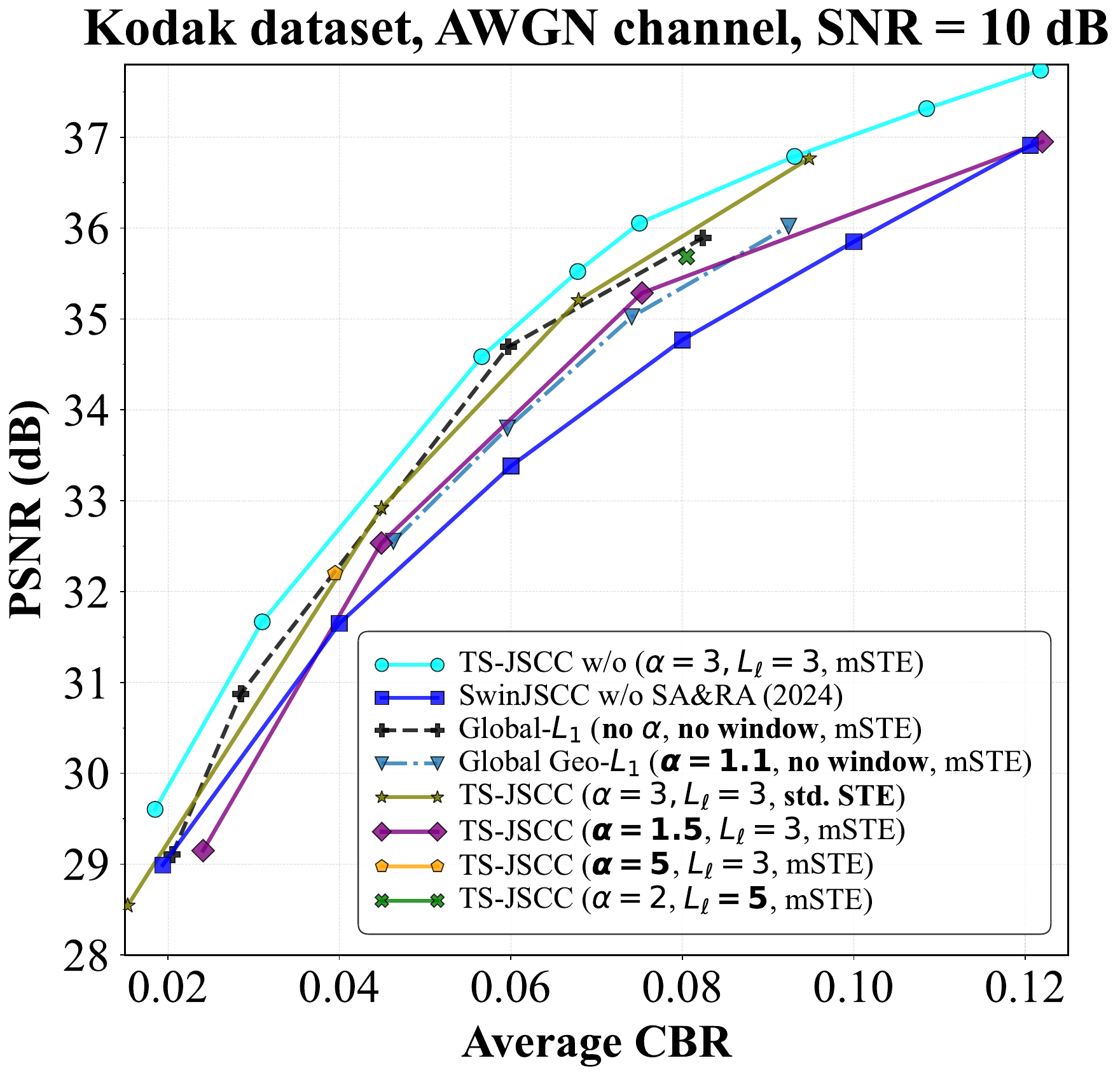}
\caption{\textbf{Ablation results on Kodak} at \(\snr=10\) dB. All TS-JSCC variants use the w/o SA\&RA setting; only the sparsification design, surrogate gradient, or hyperparameters are changed. SwinJSCC w/o SA\&RA is the same-backbone reference.}
\label{fig:ablation_awgn}
\end{figure}

Fig.~\ref{fig:ablation_awgn} first compares different sparsity structures. The Global-\(\mathrm{L1}\) curve is above the SwinJSCC w/o SA\&RA backbone over most operating points, showing that sparsity-induced resource allocation can compensate for the reconstruction loss introduced by the \(\mathrm{L1}\) regularization term. However, its active positions are scattered, so the receiver would need a dense support mask, as quantified in Table~\ref{tab:sideinfo_overhead_highres}. Global Geo-\(\mathrm{L1}\) uses exponentially increasing geometric weights along the feature-channel dimension, which can be regarded as the \(L_\ell \to \infty\) limiting case of the windowed regularizer. This produces a tail-biased support but also over-regularizes informative symbols in the active prefix, so it falls below Global-\(\mathrm{L1}\). The proposed TS-JSCC w/o SA\&RA performs best among these sparsity designs: by applying L1 pressure only around the token-specific tail boundary, it preserves useful active-prefix features while producing compact tail-contiguous support.

The other curves evaluate mSTE and the sparsification hyperparameters. Replacing mSTE with the standard STE under the same TS-JSCC w/o SA\&RA setting causes a clear performance drop, consistent with the gradient analysis in Sec.~\ref{sec:mste}. The window size and geometric factor \(\alpha\) also affect the rate--distortion trade-off, since they determine how strongly near-boundary symbols are regularized. Within the tested range, the TS-JSCC variants generally remain above the SwinJSCC w/o SA\&RA backbone, indicating that the proposed sparsification framework is effective and moderately robust to hyperparameter changes.

\section{Conclusion}

This work develops a rate/quality-adaptive and SNR-adaptive TS-JSCC framework that targets not only single-model operation over wide transmission rate and SNR ranges, but also strong reconstruction quality, low side-information overhead, and lightweight model complexity. At the resource-allocation level, the L1-based tail-structured sparsification mechanism forms prefix-active/tail-inactive latents and transmits compact boundary side information, enabling content-adaptive token-wise feature-channel allocation. At the adaptation level, lightweight stage-wise regulating modules driven by \(\lambda_{\mathrm{norm}}\) and SNR allow the adaptive model to adjust the transmission rate and adapt to different SNR conditions with marginal complexity overhead. Experiments on CIFAR-10, Kodak, and CLIC2021 under additive white Gaussian noise (AWGN) and Rayleigh fading show strong performance against the latest learned-JSCC baselines at the considered operating points, low side-information overhead, and favorable model complexity. The proposed framework also remains competitive with an idealized HPCM+LDPC+QAM separation pipeline, while retaining a simple one-shot encoder--decoder without entropy coding, context or probability prediction, explicit rate maps, or auxiliary allocation networks.

\bibliographystyle{IEEEtran}
\bibliography{refs/references}

\clearpage
\onecolumn
\setcounter{table}{0}
\renewcommand{\thetable}{S\arabic{table}}
\renewcommand{\theHtable}{S\arabic{table}}
\setcounter{equation}{0}
\renewcommand{\theequation}{S\arabic{equation}}
\renewcommand{\theHequation}{S\arabic{equation}}
\input{supplementary_content}

\end{document}

%% file: tables/tab03_sideinfo_overhead.tex
\begin{table}[t]
\centering
\caption{\textbf{Additional CBR overhead of error-free side information} in the high-resolution setting at \(\snr=10\) dB.}
\label{tab:sideinfo_overhead_highres}
\begin{tabular}{lcc}
\toprule
\textbf{Method} & \textbf{Side information} & \(\Delta\mathrm{CBR}\) \\
\midrule
Global-\(\mathrm{L1}\) & Dense mask & 0.03650 \\
HJSCC & Multi-level \(\boldsymbol{\tau}\) & 0.02000 \\
NTSCC & 16-level rate-map index & 0.00151 \\
NTSCC++ & 26-level rate-map index & 0.00188 \\
\midrule
\textbf{TS-JSCC} & \(\boldsymbol{\tau}\) in Eq.~\eqref{eq:tau} & 0.00264 \\
\textbf{TS-JSCC (16-state)} & \(\hat{\boldsymbol{\tau}}\) (4-bit index) & \textbf{0.00151} \\
\bottomrule
\end{tabular}
\end{table}

%% file: tables/tab04_complexity.tex
\begin{table}[t]
\centering
\caption{Model complexity.}
\label{tab:complexity_summary}
\begin{tabular}{lcc}
\toprule
\textbf{Method} & \textbf{Params (M)} & \textbf{FLOPs (G)} \\
\midrule
SwinJSCC w/o SA\&RA & 18.3603 & 69.3372 \\
SwinJSCC w/ RA & 23.1693 & 70.1458 \\
SwinJSCC w/ SA\&RA & 33.0339 & 71.8889 \\
HPCM source codec & 68.5053 & 120.3975 \\
HJSCC & 113.4749 & 70.1596 \\
NTSCC & 31.3585 & 131.3033 \\
NTSCC++ & 58.5944 & 84.1847 \\
\midrule
\textbf{TS-JSCC w/o SA\&RA} & 18.3603 & 69.3372 \\
\textbf{TS-JSCC (SA\&RA)} & 18.8579 & 69.3382 \\
\bottomrule
\end{tabular}
\end{table}

%% file: supplementary_content.tex
\begin{center}
{\Large\bfseries Supplementary Material for ``Single-Model Adaptive Wireless Image Transmission via Feature Sparsity Regularization''}\\[0.5em]
Xianghao Cui, Li Lan, Qi He, Bo Che, Chenyuan Feng, Zhi Chen, and Tony Q. S. Quek
\end{center}

This supplementary material provides the detailed experimental configurations underlying the learned JSCC evaluations and the separation-based baselines. It is designed to be read independently of the main manuscript: Section~I specifies the data, model, and optimization settings; Section~II gives the LDPC+QAM operating modes used for the separated transmission baselines; Section~III describes the rate-adaptive training protocol used by the adaptive TS-JSCC variants; and Section~IV details the termination-index quantization used for the 16-state TS-JSCC evaluation.

\section*{I. Experimental Configurations}

Table~S1 summarizes the configurations for the low- and high-resolution evaluations. The channel bandwidth ratio (CBR) is the number of transmitted complex channel symbols divided by the number of source scalars. Here, \(C_{\mathrm{tx}}\) denotes the maximum latent feature-channel dimension available to each token before tail sparsification, and \(\mathrm{CBR}_{\max}\) is the corresponding upper bound when all available entries are transmitted. In the high-resolution setting, \(C_{\mathrm{tx}}=192\) yields \(\mathrm{CBR}_{\max}=0.125\); the realized CBR is generally lower and varies with image content because each token transmits only its active prefix.

\input{tables/tab01_setup.tex}

The threshold \(\epsilon\) is applied to the magnitude of each complex latent symbol after sparsification; entries below this threshold are removed from the transmitted prefix. The L1-window parameters \(L_{\ell}\), \(L_r\), and \(\alpha\) determine the localized sparsity pressure around the current prefix boundary: \(L_{\ell}\) and \(L_r\) are the left and right window lengths, respectively, and \(\alpha>1\) is the geometric growth factor of the loss weight. Double power normalization is applied before thresholding and after active-prefix selection so that the sparsification criterion operates at a stable scale and the active channel-input symbols satisfy the transmit-power constraint.

\section*{II. LDPC+QAM Modes for Separation-Based Baselines}

For the BPG+LDPC+QAM and HPCM+LDPC+QAM baselines, the compressed source bitstream is transmitted using an SNR-specific LDPC+QAM mode. Table~S2 lists the mode selected for each design SNR and channel model. If \(R_{\mathrm{LDPC}}\) is the LDPC code rate and \(M\) is the QAM order, the nominal spectral efficiency is \(\eta=R_{\mathrm{LDPC}}\log_2 M\) information bits per complex channel use. The AWGN rows are also used as the fixed design modes in the SNR-mismatch experiment.

\input{tables/tab02_ldpc_qam.tex}

\section*{III. Rate-adaptive Training Protocol}

TS-JSCC (RA) uses a single model to cover multiple average-CBR operating points at a fixed SNR, whereas TS-JSCC (SA\&RA) additionally adapts the same model to the channel SNR. For either adaptive variant, a sampled multiplier \(\rho\) sets the sparsity-loss coefficient as \(\lambda_{\mathrm{L1}}=\rho\,\lambda_{\mathrm{base}}\), where \(\lambda_{\mathrm{base}}\) is the fixed base coefficient of the selected training configuration, and provides the normalized rate condition \(\lambda_{\mathrm{norm}}=\rho/\lambda_{\max}\) to the stage-wise regulating modules. Thus, the same scalar jointly controls the strength of tail sparsification during optimization and the rate condition used for feature regulation.

Rather than sampling \(\rho\) uniformly over its full numeric range, we first select an interval between two consecutive anchors and then sample \(\rho\) uniformly within that interval. This interval-based scheme prevents large-\(\rho\), high-sparsity conditions from dominating training while retaining the small-\(\rho\) conditions needed for high-CBR operation. For the high-resolution setting, the AWGN anchor sequence is [1, 4, 16, 64, 128, 256, 512, 768, 1024, 2048, 4096, 6144, 8192], with \(\lambda_{\max}=8192\), and the Rayleigh anchor sequence is [1, 64, 512, 2048, 8192, 16384, 32678], with \(\lambda_{\max}=32678\). For example, the AWGN anchors 512 and 768 define an interval from which \(\rho=640\) may be sampled, giving \(\lambda_{\mathrm{L1}}=640\,\lambda_{\mathrm{base}}\) and \(\lambda_{\mathrm{norm}}=640/8192\approx0.078\).

\section*{IV. Termination-Index Quantization for the 16-State Evaluation}

The high-resolution TS-JSCC w/o SA\&RA latent has a maximum active prefix length of 96 complex symbols per token. The full-precision transmission rule therefore conveys a termination index \(\tau_s\in\{0,\ldots,96\}\) for each token, requiring seven bits per token. To evaluate a lower-overhead variant without retraining, we use the 16-state codebook
\begin{equation}
\mathcal{Q}_{16}=\{0,4,6,8,10,12,16,20,24,28,36,44,52,60,72,96\}.
\end{equation}
After complex-symbol magnitude thresholding determines \(\tau_s\), the transmitter replaces it by the nearest codebook value,
\begin{equation}
\hat{\tau}_s=\operatorname*{arg\,min}_{q\in\mathcal{Q}_{16}}\lvert\tau_s-q\rvert.
\end{equation}
If two codebook values are equally close to \(\tau_s\), the shorter prefix is selected. The 16-state alphabet requires 4 bits per token. The payload-CBR curves in Fig.~6 of the main manuscript exclude this small side-information term; its capacity-equivalent value is reported separately in Table~I.

We evaluate this option at 10 dB complex AWGN on Kodak and CLIC2021 using the same eight TS-JSCC w/o SA\&RA checkpoints as the full-precision curves and the same image preprocessing; no fine-tuning is performed. Table~S3 reports the difference relative to the full-precision termination-index rule at matched checkpoints. The 16-state quantization retains nearly identical reconstruction quality on both datasets.

\begin{table}[H]
\centering
\caption{16-state termination-index quantization versus full precision at 10 dB AWGN. Differences are in PSNR (dB) at matched checkpoints.}
\label{tab:q16_results}
\begin{tabular}{lcc}
\toprule
\textbf{Dataset} & \textbf{Mean difference} & \textbf{Worst difference} \\
\midrule
Kodak & \(-0.047\) & \(-0.217\) \\
CLIC2021 & \(-0.082\) & \(-0.432\) \\
\bottomrule
\end{tabular}
\end{table}

%% file: tables/tab01_setup.tex
\begin{table}[H]
\centering
\footnotesize
\setlength{\tabcolsep}{4.2pt}
\renewcommand{\arraystretch}{1.06}
\caption{Detailed experimental configurations for the learned JSCC evaluations. \(C_{\mathrm{tx}}\) is the maximum latent feature-channel dimension available to a token before tail sparsification, and \(\mathrm{CBR}_{\max}\) is the corresponding upper bound on CBR.}
\label{tab:setup_summary}
\begin{tabular}{@{}>{\raggedright\arraybackslash}m{4.5cm}>{\centering\arraybackslash}m{4.8cm}>{\centering\arraybackslash}m{5.8cm}@{}}
\toprule
\textbf{Item} & \textbf{Low-resolution setting} & \textbf{High-resolution setting} \\
\midrule
Train / test data & CIFAR-10 / CIFAR-10 & DIV2K / Kodak, CLIC2021 \\
Backbone depths & \([2,4]\) & \([2,2,6,2]\) \\
Backbone channels & \([128,256]\) & \([128,192,256,320]\) \\
Backbone attention heads & \([4,8]\) & \([4,6,8,10]\) \\
Backbone window size & \(2\) & \(8\) \\
Max. selectable channels \(C_{\mathrm{tx}}\) & \(96\;(\mathrm{CBR}_{\max}=1)\) & \(192\;(\mathrm{CBR}_{\max}=0.125)\) \\
Threshold \(\epsilon\) & \(10^{-2}\) & \(10^{-2}\) \\
Main AWGN L1-window setting & \(L_{\ell}=3,\;L_r=1,\;\alpha=3\) & \(L_{\ell}=3,\;L_r=1,\;\alpha=3\) \\
Double power normalization & Yes & Yes \\
Optimizer / learning rate & Adam / \(10^{-4}\) & Adam / \(10^{-4}\) \\
Batch size & \(128\) & \(16\) \\
\bottomrule
\end{tabular}
\end{table}

%% file: tables/tab02_ldpc_qam.tex
\begin{table}[H]
\centering
\scriptsize
\setlength{\tabcolsep}{3.0pt}
\renewcommand{\arraystretch}{1.06}
\caption{LDPC+QAM modes for the separation-based baselines. The design SNR determines the selected LDPC code rate and QAM order; \(\eta=R_{\mathrm{LDPC}}\log_2 M\) is the nominal spectral efficiency in information bits per complex channel use. The AWGN rows are also used as fixed design modes in the SNR-mismatch experiment.}
\label{tab:ldpc_qam_schedule}
\begin{tabular}{llccc}
\toprule
\textbf{Channel} & \textbf{SNR/design (dB)} & \textbf{LDPC rate} & \textbf{Modulation} & \(\boldsymbol{\eta}\) \\
\midrule
\multirow{5}{*}{AWGN}
& 1  & \(1/3\) & 4-QAM  & \(2/3\) \\
& 4  & \(1/2\) & 4-QAM  & \(1\) \\
& 7  & \(2/3\) & 4-QAM  & \(4/3\) \\
& 10 & \(2/3\) & 16-QAM & \(8/3\) \\
& 13 & \(1/2\) & 64-QAM & \(3\) \\
\midrule
\multirow{4}{*}{Rayleigh}
& 3  & \(1/3\) & 4-QAM & \(2/3\) \\
& 5  & \(1/2\) & 4-QAM & \(1\) \\
& 8  & \(2/3\) & 4-QAM & \(4/3\) \\
& 11 & \(3/4\) & 4-QAM & \(3/2\) \\
\bottomrule
\end{tabular}
\end{table}

%% file: refs/references.bib
@ARTICLE{Shannon1948,
  author={Shannon, C. E.},
  journal={The Bell System Technical Journal}, 
  title={A mathematical theory of communication}, 
  year={1948},
  volume={27},
  number={3},
  pages={379-423},
  doi={10.1002/j.1538-7305.1948.tb01338.x}}

@article{Shannon1959,
  title={Coding theorems for a discrete source with a fidelity criterion},
  author={Shannon, Claude E and others},
  journal={IRE Nat. Conv. Rec},
  volume={4},
  number={142-163},
  pages={1},
  year={1959}
}

@book{CoverThomas2006,
  title={Elements of information theory},
  author={Cover, Thomas M},
  year={1999},
  publisher={John Wiley \& Sons}
}

@article{Wallace1992,
author = {Wallace, Gregory K.},
title = {The JPEG still picture compression standard},
year = {1991},
issue_date = {April 1991},
publisher = {Association for Computing Machinery},
address = {New York, NY, USA},
volume = {34},
number = {4},
issn = {0001-0782},
doi = {10.1145/103085.103089},
journal = {Commun. ACM},
month = apr,
pages = {30–44},
numpages = {15}
}

@ARTICLE{RichardsonUrbanke2001,
  author={Richardson, T.J. and Urbanke, R.L.},
  journal={IEEE Transactions on Information Theory}, 
  title={The capacity of low-density parity-check codes under message-passing decoding}, 
  year={2001},
  volume={47},
  number={2},
  pages={599-618},
  doi={10.1109/18.910577}}

@ARTICLE{Arikan2009,
  author={Arikan, Erdal},
  journal={IEEE Transactions on Information Theory}, 
  title={Channel Polarization: A Method for Constructing Capacity-Achieving Codes for Symmetric Binary-Input Memoryless Channels}, 
  year={2009},
  volume={55},
  number={7},
  pages={3051-3073},
  doi={10.1109/TIT.2009.2021379}}

@ARTICLE{Bourtsoulatze2019,
  author={Bourtsoulatze, Eirina and Burth Kurka, David and Gündüz, Deniz},
  journal={IEEE Transactions on Cognitive Communications and Networking}, 
  title={Deep Joint Source-Channel Coding for Wireless Image Transmission}, 
  year={2019},
  volume={5},
  number={3},
  pages={567-579},

  doi={10.1109/TCCN.2019.2919300}}

@InProceedings{Liu2021Swin,
    author    = {Liu, Ze and Lin, Yutong and Cao, Yue and Hu, Han and Wei, Yixuan and Zhang, Zheng and Lin, Stephen and Guo, Baining},
    title     = {Swin Transformer: Hierarchical Vision Transformer Using Shifted Windows},
    booktitle = {Proceedings of the IEEE/CVF International Conference on Computer Vision (ICCV)},
    month     = {October},
    year      = {2021},
    pages     = {10012-10022}
}

@ARTICLE{SwinJSCC,
  author={Yang, Ke and Wang, Sixian and Dai, Jincheng and Qin, Xiaoqi and Niu, Kai and Zhang, Ping},
  journal={IEEE Transactions on Cognitive Communications and Networking}, 
  title={SwinJSCC: Taming Swin Transformer for Deep Joint Source-Channel Coding}, 
  year={2025},
  volume={11},
  number={1},
  pages={90-104},
  doi={10.1109/TCCN.2024.3424842}}

@ARTICLE{Dai2022NTSCC,
  author={Dai, Jincheng and Wang, Sixian and Tan, Kailin and Si, Zhongwei and Qin, Xiaoqi and Niu, Kai and Zhang, Ping},
  journal={IEEE Journal on Selected Areas in Communications},
  title={Nonlinear Transform Source-Channel Coding for Semantic Communications},
  year={2022},
  volume={40},
  number={8},
  pages={2300-2316},
  doi={10.1109/JSAC.2022.3180802}}

@ARTICLE{Wang2023ImprovedNTSCC,
  author={Wang, Sixian and Dai, Jincheng and Qin, Xiaoqi and Si, Zhongwei and Niu, Kai and Zhang, Ping},
  journal={IEEE Journal of Selected Topics in Signal Processing},
  title={Improved Nonlinear Transform Source-Channel Coding to Catalyze Semantic Communications},
  year={2023},
  volume={17},
  number={5},
  pages={1022-1037},
  doi={10.1109/JSTSP.2023.3304140}}

@INPROCEEDINGS{Bian2023DeepJSCCLpp,
  author={Bian, Chenghong and Shao, Yulin and Gündüz, Deniz},
  booktitle={Proc. IEEE Global Communications Conference (GLOBECOM)},
  title={DeepJSCC-l++: Robust and Bandwidth-Adaptive Wireless Image Transmission},
  year={2023},
  pages={3148--3154},
  doi={10.1109/GLOBECOM54140.2023.10436878}}

@ARTICLE{Li2025STARJSCC,
  author={Li, Xiangcheng and Ban, Dongri and Ruan, Zhaokai and Yue, Xiuyu and Chen, Haiqiang and Sun, Youming},
  journal={Scientific Reports},
  title={A Star Modulation Network for Wireless Image Semantic Transmission},
  year={2025},
  volume={15},
  pages={31127},
  doi={10.1038/s41598-025-16753-4}}

@ARTICLE{Xu2022ADJSCC,
  author={Xu, Jialong and Ai, Bo and Chen, Wei and Yang, Ang and Sun, Peng and Rodrigues, Miguel},
  journal={IEEE Transactions on Circuits and Systems for Video Technology},
  title={Wireless Image Transmission Using Deep Source Channel Coding With Attention Modules},
  year={2022},
  volume={32},
  number={4},
  pages={2315-2328},
  doi={10.1109/TCSVT.2021.3082521}}

@Article{RateAdaptiveJSCC_Song2023,
AUTHOR = {Song, Mengshu and Ma, Nan and Dong, Chen and Xu, Xiaodong and Zhang, Ping},
TITLE = {Deep Joint Source-Channel Coding for Wireless Image Transmission with Adaptive Models},
JOURNAL = {Electronics},
VOLUME = {12},
YEAR = {2023},
NUMBER = {22},
ARTICLE-NUMBER = {4637},
ISSN = {2079-9292},
DOI = {10.3390/electronics12224637}
}

@article{RateAdaptiveJSCC_Zhang2023,
  title={Predictive and adaptive deep coding for wireless image transmission in semantic communication},
  author={Zhang, Wenyu and Zhang, Haijun and Ma, Hui and Shao, Hua and Wang, Ning and Leung, Victor CM},
  journal={IEEE Transactions on Wireless Communications},
  volume={22},
  number={8},
  pages={5486--5501},
  year={2023},
  publisher={IEEE}
}

@INPROCEEDINGS{RateAdaptiveJSCC_YangKim2022,
  author={Yang, Mingyu and Kim, Hun-Seok},
  booktitle={ICASSP 2022 - 2022 IEEE International Conference on Acoustics, Speech and Signal Processing (ICASSP)}, 
  title={Deep Joint Source-Channel Coding for Wireless Image Transmission with Adaptive Rate Control}, 
  year={2022},
  volume={},
  number={},
  pages={5193-5197},
  doi={10.1109/ICASSP43922.2022.9746335}}

@article{Tibshirani1996,
  title={Regression shrinkage and selection via the lasso},
  author={Tibshirani, Robert},
  journal={Journal of the Royal Statistical Society Series B: Statistical Methodology},
  volume={58},
  number={1},
  pages={267--288},
  year={1996},
  publisher={Oxford University Press}
}

@InProceedings{DIV2K,
author = {Agustsson, Eirikur and Timofte, Radu},
title = {NTIRE 2017 Challenge on Single Image Super-Resolution: Dataset and Study},
booktitle = {Proceedings of the IEEE Conference on Computer Vision and Pattern Recognition (CVPR) Workshops},
month = {July},
year = {2017}
}

@article{CLIC2021,
  title={Clic 2020: Challenge on learned image compression},
  author={Toderici, George and Theis, Lucas and Johnston, Nick and Agustsson, Eirikur and Mentzer, Fabian and Ball{\'e}, Johannes and Shi, Wenzhe and Timofte, Radu},
  journal={Retrieved March},
  volume={29},
  pages={2021},
  year={2020}
}

@misc{CIFAR10,
  title={Learning multiple layers of features from tiny images},
  author={Krizhevsky, Alex and Hinton, Geoffrey and others},
  howpublished={Technical report, University of Toronto},
  year={2009},
  note={Toronto, ON, Canada}
}

@misc{Kodak,
  title        = {Kodak PhotoCD Dataset},
  year         = {1993},
  howpublished = {Online},
  url          = {http://r0k.us/graphics/kodak/},
}

@article{wang2015mcscast,
  title={A wireless video multicasting scheme based on multi-scale compressed sensing},
  author={Wang, Anhong and Wu, Qingdian and Ma, Xiaoli and Zeng, Bing},
  journal={EURASIP Journal on Advances in Signal Processing},
  volume={2015},
  number={1},
  pages={74},
  year={2015},
  doi     = {10.1186/s13634-015-0258-2},
  publisher={Springer}
}

@inproceedings{jakubczak2010softcast,
  title={SoftCast: One-size-fits-all wireless video},
  author={Jakubczak, Szymon and Katabi, Dina},
  booktitle={Proceedings of the ACM SIGCOMM 2010 conference},
  pages={449--450},
  year={2010}
}

@article{Zhang2025HJSCC,
title={Learned Image Transmission with Hierarchical Variational Autoencoder},
journal={Proceedings of the AAAI Conference on Artificial Intelligence},
volume={39},
DOI={10.1609/aaai.v39i12.33442},
author={Zhang, Guangyi and Li, Hanlei and Cai, Yunlong and Hu, Qiyu and Yu, Guanding and Zhang, Runmin}, 
year={2025},
month={Apr.},
pages={13215-13223} }

@INPROCEEDINGS{Chen2023EntropyAware,
  author={Chen, Weixuan and Chen, Yuhao and Yang, Qianqian and Huang, Chongwen and Wang, Qian and Zhang,Zhaoyang},
  booktitle={GLOBECOM 2023 - 2023 IEEE Global Communications Conference}, 
  title={Deep Joint Source-Channel Coding for Wireless Image Transmission with Entropy-Aware Adaptive Rate Control}, 
  year={2023},
  volume={},
  number={},
  pages={2239-2244},
  doi={10.1109/GLOBECOM54140.2023.10437482}}

@ARTICLE{Zhang2023VLSCC,
  author={Zhang, Bowen and Qin, Zhijin and Li, Geoffrey Ye},
  journal={IEEE Journal of Selected Topics in Signal Processing}, 
  title={Semantic Communications With Variable-Length Coding for Extended Reality}, 
  year={2023},
  volume={17},
  number={5},
  pages={1038-1051},
  doi={10.1109/JSTSP.2023.3300509}}

@article{Bengio2013STE,
  title={Estimating or propagating gradients through stochastic neurons for conditional computation},
  author={Bengio, Yoshua and L{\'e}onard, Nicholas and Courville, Aaron},
  journal={arXiv preprint arXiv:1308.3432},
  year={2013}
}

@inproceedings{WenSSL,
 author = {Wen, Wei and Wu, Chunpeng and Wang, Yandan and Chen, Yiran and Li, Hai},
 booktitle = {Advances in Neural Information Processing Systems},
 editor = {D. Lee and M. Sugiyama and U. Luxburg and I. Guyon and R. Garnett},
 pages = {},
 publisher = {Curran Associates, Inc.},
 title = {Learning Structured Sparsity in Deep Neural Networks},
 url = {https://proceedings.neurips.cc/paper_files/paper/2016/file/41bfd20a38bb1b0bec75acf0845530a7-Paper.pdf},
 volume = {29},
 year = {2016}
}

@InProceedings{LiuNetworkSlimming,
author = {Liu, Zhuang and Li, Jianguo and Shen, Zhiqiang and Huang, Gao and Yan, Shoumeng and Zhang, Changshui},
title = {Learning Efficient Convolutional Networks Through Network Slimming},
booktitle = {Proceedings of the IEEE International Conference on Computer Vision (ICCV)},
month = {Oct},
year = {2017}
}

@InProceedings{HeChannelPruning,
author = {He, Yihui and Zhang, Xiangyu and Sun, Jian},
title = {Channel Pruning for Accelerating Very Deep Neural Networks},
booktitle = {Proceedings of the IEEE International Conference on Computer Vision (ICCV)},
month = {Oct},
year = {2017}
}

@article{Han2015Pruning,
  title={Deep compression: Compressing deep neural networks with pruning, trained quantization and huffman coding},
  author={Han, Song and Mao, Huizi and Dally, William J},
  journal={arXiv preprint arXiv:1510.00149},
  year={2015}
}

@inproceedings{Jang2017GumbelSoftmax,
  author = {Jang, Eric and Gu, Shixiang and Poole, Ben},
  title = {Categorical Reparameterization with Gumbel-Softmax},
  booktitle = {Proceedings of the International Conference on Learning Representations (ICLR)},
  year = {2017}
}

@inproceedings{Maddison2017Concrete,
  author = {Maddison, Chris J. and Mnih, Andriy and Teh, Yee Whye},
  title = {The Concrete Distribution: A Continuous Relaxation of Discrete Random Variables},
  booktitle = {Proceedings of the International Conference on Learning Representations (ICLR)},
  year = {2017}
}

@article{YuanLin2006GroupLasso,
  author = {Yuan, Ming and Lin, Yi},
  title = {Model Selection and Estimation in Regression with Grouped Variables},
  journal = {Journal of the Royal Statistical Society: Series B},
  volume = {68},
  number = {1},
  pages = {49--67},
  year = {2006}
}

@inproceedings{Louizos2018L0,
  author = {Louizos, Christos and Welling, Max and Kingma, Diederik P.},
  title = {Learning Sparse Neural Networks through $L_0$ Regularization},
  booktitle = {Proceedings of the International Conference on Learning Representations (ICLR)},
  year = {2018}
}

@article{Molchanov2017VariationalDropout,
  author = {Molchanov, Dmitry and Ashukha, Arsenii and Vetrov, Dmitry},
  title = {Variational Dropout Sparsifies Deep Neural Networks},
  journal = {Proceedings of Machine Learning Research},
  volume = {70},
  pages = {2498--2507},
  year = {2017}
}

@article{Yin2019UnderstandingSTE,
  author = {Yin, Penghang and Lyu, Jiancheng and Zhang, Shuai and Osher, Stanley and Qi, Yingyong and Xin, Jack},
  title = {Understanding Straight-Through Estimator in Training Activation Quantized Neural Nets},
  journal = {arXiv preprint arXiv:1903.05662},
  year = {2019}
}

@misc{Bellard2014,
  author       = {Fabrice Bellard},
  title        = {BPG Image format},
  year         = {2014},
  howpublished = {\url{https://bellard.org/bpg/}},
  note         = {Accessed: 2026-03-04}
}

@inproceedings{He2022ELIC,
  author    = {He, Dailan and Yang, Ziming and Peng, Weikun and Ma, Rui and Qin, Hongwei and Wang, Yan},
  title     = {{ELIC}: Efficient Learned Image Compression With Unevenly Grouped Space-Channel Contextual Adaptive Coding},
  booktitle = {Proceedings of the IEEE/CVF Conference on Computer Vision and Pattern Recognition},
  pages     = {5718--5727},
  year      = {2022}
}

@inproceedings{HPCM2025,
  author    = {Li, Yuqi and Zhang, Haotian and Li, Li and Liu, Dong},
  title     = {Learned Image Compression With Hierarchical Progressive Context Modeling},
  booktitle = {Proceedings of the IEEE/CVF International Conference on Computer Vision},
  pages     = {18834--18843},
  year      = {2025}
}

@inproceedings{DCAE2025,
  author    = {Lu, Jingbo and Zhang, Leheng and Zhou, Xingyu and Li, Mu and Li, Wen and Gu, Shuhang},
  title     = {Learned Image Compression With Dictionary-Based Entropy Model},
  booktitle = {Proceedings of the IEEE/CVF Conference on Computer Vision and Pattern Recognition},
  pages     = {12850--12859},
  year      = {2025}
}

@inproceedings{LALIC2025,
  author    = {Feng, Donghui and Cheng, Zhengxue and Wang, Shen and Wu, Ronghua and Hu, Hongwei and Lu, Guo and Song, Li},
  title     = {Linear Attention Modeling for Learned Image Compression},
  booktitle = {Proceedings of the IEEE/CVF Conference on Computer Vision and Pattern Recognition},
  pages     = {7623--7632},
  year      = {2025}
}

@inproceedings{MambaIC2025,
  author    = {Zeng, Fanhu and Tang, Hao and Shao, Yihua and Chen, Siyu and Shao, Ling and Wang, Yan},
  title     = {{MambaIC}: State Space Models for High-Performance Learned Image Compression},
  booktitle = {Proceedings of the IEEE/CVF Conference on Computer Vision and Pattern Recognition},
  pages     = {18041--18050},
  year      = {2025}
}
